\documentclass[sigconf]{acmart}

\copyrightyear{2026}
\acmYear{2026}
\setcopyright{cc}
\setcctype{by}
\acmConference[WWW '26]{Proceedings of the ACM Web Conference 2026}{April 13--17, 2026}{Dubai, United Arab Emirates}
\acmBooktitle{Proceedings of the ACM Web Conference 2026 (WWW '26), April 13--17, 2026, Dubai, United Arab Emirates}
\acmPrice{}
\acmDOI{10.1145/3774904.3792476}
\acmISBN{979-8-4007-2307-0/2026/04}

\usepackage{algorithm}
\usepackage{algorithmic}   % replaces algpseudocode

\usepackage{makecell}
\newcolumntype{Y}{>{\raggedright\arraybackslash}X}
\newtheorem{lemma}{Lemma}
\newtheorem{proposition}{Proposition}
\newtheorem{theorem}{Theorem}

\usepackage{enumitem}
\begin{document}

\title{Reliable Non-Leveled Homomorphic Encryption for Web Services}

\author{Baigang Chen}
\orcid{0009-0009-4928-5362}
\affiliation{
  \institution{University of Minnesota}
  \city{Twin Cities}
  \state{MN}
  \country{United States}}
\email{chen9464@umn.edu}
\authornote{Most of this work was carried out at the University of Washington, Seattle.}
\author{Dongfang Zhao}
\orcid{0000-0002-0677-634X}
\affiliation{
  \institution{University of Washington}
  \city{Tacoma}
  \state{WA}
  \country{United States}}
\email{dzhao@cs.washington.edu}

\begin{abstract}
With the ubiquitous deployment of web services, ensuring data confidentiality has become a challenging imperative. Fully Homomorphic Encryption (FHE) presents a powerful solution for processing encrypted data; however, its widespread adoption is severely constrained by two fundamental bottlenecks: substantial computational overhead and the absence of a built-in automatic error correction mechanism. These limitations render the deployment of FHE in real-world, complex network environments impractical.

To address this dual challenge, this work puts forward a new FHE framework that enhances computational efficiency and integrates an automatic error correction capability through new encoding techniques and an algebraic reliability layer.Our prototype is evaluated through encrypted low-degree activation timing, one experimental public Refresh skeleton invocation, and transport-fault simulations for the Ring--BCH layer. Our current prototype quantifies the cost of encrypted low-degree activation
evaluation, the additional latency of an experimental public Refresh skeleton,
and the robustness gained from the Ring--BCH transport layer. The Refresh
prototype should be interpreted as a skeleton rather than a complete CKKS
bootstrapping implementation, since it uses a low-degree surrogate rather than
a validated EvalMod circuit. In transport-fault simulations, the BCH interleaver
reduces failure rates to below $0.5\%$ under bursty faults and keeps the modeled
accuracy within $0.5$ percentage points of the plaintext baseline.
\end{abstract}
% --- CCS Concepts (use the XML block + \ccsdesc lines) ---



% --- Keywords ---

\maketitle

\section{Introduction}

Ensuring data privacy is a central challenge in modern web services, particularly for data-intensive applications like machine learning~\cite{larch2019currency}. Fully Homomorphic Encryption (FHE) provides a powerful cryptographic foundation for such services. The CKKS scheme~\cite{cheon2017homomorphic}, which supports arithmetic on encrypted real numbers, is especially well-suited for machine learning workloads. However, two significant obstacles hinder the practical and dependable deployment of CKKS in real-world services. The first is the rigidity of the conventional leveled execution paradigm, which requires developers to predict the computation's multiplicative depth in advance. The second is the fragility of ciphertexts, which can be corrupted during transport across distributed system components, degrading accuracy and reliability.

These challenges manifest as concrete engineering and performance problems. In a standard leveled CKKS workflow, noise from multiplications is managed by descending a pre-selected chain of moduli via a \texttt{Rescale} operation. This forces a trade-off: developers must either overestimate the required multiplicative depth, leading to inefficiently large parameters, or risk computation failure if the depth is underestimated. This upfront depth-planning requirement complicates development and can lead to unnecessary latency. Concurrently, in realistic deployment settings where ciphertexts traverse clients, servers, and storage systems, even infrequent bit-flips from network or hardware faults can cause complete decryption failures or silent numerical errors~\cite{GlitchFHE25}. Conventional integrity mechanisms like hash-based retries are often unsuitable, as the additional round-trips they require can substantially inflate the tail latency of the service~\cite{sriraman2017tail}.

This paper presents a new FHE framework designed to address these dual challenges of usability and reliability. Our first component is a non-leveled variant of CKKS that operates at a fixed modulus. It replaces the standard \texttt{Rescale} operation with a  \texttt{Refresh} operation, which periodically resets the ciphertext's scale to maintain numerical precision. This approach removes the need to plan computational depth, simplifying parameter selection and service deployment. Our second component is a lightweight error-correction layer integrated with a permutation-based interleaver. This mechanism effectively disperses bursty transmission errors across the ciphertext structure, enabling their correction with only millisecond-level overhead while remaining fully compatible with homomorphic operations.

We implemented a \textsc{HElib}-based prototype and evaluated two aspects of
the design: encrypted low-degree activation evaluation with an experimental
public Refresh skeleton, and transport-fault robustness from the Ring--BCH
interleaving layer. The Refresh experiment measures the cost of the proposed
homomorphic decrypt-and-reencrypt path, but does not yet constitute a complete
CKKS bootstrapping implementation. Separately, the Ring--BCH layer demonstrates
strong robustness under bursty channel faults, reducing failure rates to below
$0.5\%$ and keeping modeled accuracy within $0.5$ percentage points of the
plaintext baseline.

\paragraph{Contributions:}
\begin{itemize}[leftmargin=*, nosep]
  \item[\textbf{1.}] We present a fixed-modulus CKKS workflow that uses a periodic Refresh operation in place of leveled \texttt{Rescale}. This eliminates the need for multiplicative depth planning, simplifying the parameterization and deployment of FHE-based web services.

  \item[\textbf{2.}] We design a code-based reliability layer combined with a permutation interleaver. This layer disperses and corrects burst errors that occur during data transmission with low runtime overhead, enhancing the robustness of FHE in web services.

\item[\textbf{3.}] We provide a \textsc{HElib}-based prototype evaluation that
measures encrypted cubic activation latency, the additional cost of one
experimental Refresh skeleton invocation, and the robustness of the Ring--BCH
transport layer under bursty faults.
\end{itemize}

\section{Related Work}
In this section, we review the developing history of FHE and the current works that bridge the gap between theoretical foundations and real-life deployments. 
\subsection{Fully Homomorphic Encryption}

Early fully homomorphic encryption (FHE) enabled exact computation over modular rings with noise controlled by modulus switching or bootstrapping. Gentry’s construction~\cite{gentry2009fully} led to leveled schemes such as BGV~\cite{brakerski2012fg}, BFV~\cite{fan2012somewhat}, and DGHV~\cite{van2010fully}. TFHE~\cite{chillotti2020tfhe} advanced bit-level computation with fast gate bootstrapping. Recent surveys provide implementation guidance and trade-offs for practical deployments~\cite{gong2024practical,ZhaoCompileTime25,ZhaoSimilarity25}. These designs are strong for exact integer logic but map poorly to the real-valued kernels that dominate privacy-preserving machine learning (PPML) on the web.

CKKS~\cite{cheon2017homomorphic} addresses PPML by supporting approximate arithmetic and SIMD packing, enabling linear algebra over encrypted feature vectors and model weights. CKKS underpins many PPML pipelines in inference and analytics, yet three issues persist in practice: rapid noise growth that forces careful scale management, rescaling chains that complicate parameter tuning, and bootstrapping costs that limit depth and throughput. In distributed settings, ciphertext transmission adds another failure mode: rare bit flips can derail decryption and downstream ML tasks. Our work targets these PPML pain points by removing the leveled schedule and adding an algebraic reliability layer that fits the adjusted CKKS ring, improving robustness while preserving the programming model.

\subsection{FHE System Optimization}
Recent efforts have been focusing on improving the practical usability and performance of CKKS through software libraries and hardware acceleration. Libraries such as Microsoft SEAL~\cite{fawaz2021comparative}, PALISADE~\cite{takeshita2025heprofiler}, and Lattigo~\cite{mouchet2020lattigo} offer high-level CKKS APIs alongside optimized FFT back-ends, while GPU-accelerated solutions~\cite{wang2023hebooster,zhao2024survey,CletZuber21,rohloff2014scalable} provide deployment strategies tailored for commercial cloud platforms. Advances in CKKS optimization in low-complexity ciphertext multiplication~\cite{AkheratiZhang23, chillotti2016faster} and client-efficient lightweight variants~\cite{CheonLightweightCKKS25} further reduce the theoretical complexity. Concurrently, coded-BKW techniques, combining the Blum–Kalai–Wasserman algorithm with algebraic error-correcting codes, have been investigated to improve decryption robustness~\cite{maartensson2019asymptotic}. However, beyond those approaches, system-level implementations~\cite{mazzone2025efficient} that focus on efficient computation still involve significant parameter inflation and remain limited to integer-based FHE schemes. Our work presents a unified BCH-coded framework based on the CKKS scheme that supports approximate-number arithmetic, achieves strictly bounded multiplicative noise growth, and guarantees deterministic, bit-exact decryption without compromising IND-CPA security.

Moreover, to address storage and bandwidth limitations in practical deployments, several advanced compression and key management techniques have been proposed. Additive-HE helper compression reduces ciphertext sizes in LWE-based schemes by exploiting homomorphic aggregation~\cite{HE2023}. Parallel-caching frameworks such as PawCache~\cite{tawose2023toward} and SILCA~\cite{ZhaoSilca23} precompute rotation and facilitate relinearization keys in radix form, significantly reducing key-generation latency and facilitating efficient batch evaluation. These system-level optimizations represent promising directions for integration into the non-leveled CKKS framework to enhance its performance and deployability.
\subsection{FHE Bootstrapping and Fault Tolerance}
On the bootstrapping front, polynomial-switching techniques with multi-digit decomposition achieve sub-second refresh times~\cite{lee2025enhanced}, while Bit-wise and small-integer bootstrapping variants~\cite{bae2024bits,bae2024ints,TanHighPrecision25} have demonstrated the feasibility of exact arithmetic over small domains at competitive speeds. Bootstrapping for binary input and non-sparse keys are efficient with modulus raising techniques~\cite{BossuatBootstrapping20, CheonLightweightCKKS25,cheon2018bootstrapping}. More recent parallelized bootstrapping approaches~\cite{cheon2025ship} leverage multi-core and SIMD architectures to further reduce latency. At the protocol level, CKKS has been adopted in privacy-preserving analytics, secure multi-party computation, and end-to-end encrypted machine learning pipelines~\cite{HanHEaaNNB24}; its IND-CPA security model has been rigorously formalized and extended to withstand adaptive adversaries~\cite{hwang2025ckks}.

In the quantum domain, early fault-tolerant FHE constructions employ CSS stabilizer codes to preserve logical qubit integrity during arbitrary quantum computations. Liang and Yang’s QFHE scheme~\cite{fault_tor2015} uses transversal Clifford gates and magic-state injection to maintain fault tolerance under encryption, while later refinements unify encoding and encryption within a single CSS layer, thereby improving both security levels and correctable-fault threshold~\cite{sohn2024error}. 

To the best of our knowledge, no prior work combines the approximate-arithmetic capabilities of CKKS with an algebraic outer error-correcting code that remains closed under the scheme’s ring operations. We close this gap by (i) proposing a \emph{non-leveled} CKKS variant that operates at a fixed scale; and (ii) integrating a Hensel-lifted BCH ideal, together with automorphism-based interleaving, so that codewords are preserved under addition and multiplication. This composition enables deterministic, bit-exact recovery without degrading IND-CPA security.

\section{Methodology}
In this section, we describe our core construction and analysis.
\subsection{Building Blocks}
All logarithms are base 2 unless stated otherwise. For a real number $r$, let $\lfloor r \rceil$ be the nearest integer, rounding up in case of a tie. For an integer $q$, we identify $\mathbb{Z}_q$ with $\mathbb{Z} \cap (-q/2, q/2]$, and write $[z]_q$ for the reduction of $z$ into that interval. 
\subsubsection{Cyclotomic Ring and Canonical Embedding}
Let $M\in\mathbb{N}$ and $\Phi_M(X)$ be the $M$-th cyclotomic polynomial of degree $N=\varphi(M)$. We work over $R=\mathbb{Z}[X]/(\Phi_M)$ and its residue ring $R_q=R/qR$. Write $a\in R$ as $a(X)=\sum_{j=0}^{N-1}a_jX^j$ and identify it with its coefficient vector $(a_0,\ldots,a_{N-1})\in\mathbb{R}^N$; we use the usual coefficient $\|\cdot\|_\infty$ and $\|\cdot\|_1$ norms. The canonical embedding $\sigma(a)$ evaluates $a$ at the primitive $M$-th roots of unity: $\sigma(a)=(a(\zeta_M^j))_{j\in\mathbb{Z}_M^*}\in\mathbb{C}^N$ with $\zeta_M=e^{-2\pi i/M}$, and we set $\|a\|_{\mathrm{can},\infty}:=\|\sigma(a)\|_\infty$; note $\|\sigma\|_{\mathrm{op}}\le\sqrt{N}$. The image of $\sigma$ lies in the real Hermitian subspace $\{z\in\mathbb{C}^N: z_{-j}=\overline{z_j}\}$, which (via a unitary change of basis) is isometric to $\mathbb{R}^N$~\cite{shimura1963arithmetic}.

\subsubsection{Gaussian Sampling and RLWE}
Let $H$ be the Hermitian space from above. For $r>0$, write the (spherical) Gaussian on $H$ as $\mathcal{D}_H(0,r^2 I)$ with density proportional to $\exp(-\pi\|z\|^2/r^2)$. For an elliptical variant with per-coordinate scales $r_1,\ldots,r_N>0$, use $\mathcal{D}_H\!\big(0,\mathrm{diag}(r_1^2,\ldots,r_N^2)\big)$. If $U:\mathbb{R}^N\!\to H$ is a fixed real isometry, then sampling $z\!\leftarrow\!\mathcal{N}(0,\mathrm{diag}(r_1^2,\ldots,r_N^2))$ and outputting $U z$ realizes this distribution. Mapping via $\mathrm{CRT}_M^{-1}\!\circ U$ gives a continuous distribution $\Psi_r$ over $R\otimes\mathbb{R}\cong\mathbb{R}[X]/(\Phi_M)$. Discretizing to the dual lattice by rounding yields the discrete Gaussian $\chi:=\lfloor \Psi_r\rfloor_{R^\vee}$, which we use as the error distribution.

\begin{definition}[RLWE distribution]
Let $q\ge2$, $R_q:=R/qR$, and $R_q^\vee:=R^\vee/qR^\vee$. For a secret $s\in R_q^\vee$ and error $\chi$, define $\mathsf{A}_{q,\chi}(s)$ over $R_q\times R_q^\vee$ by: sample $a\!\leftarrow\!R_q$ uniformly and $e\!\leftarrow\!\chi$, then return $(a,\;a\cdot s+e)$.
\end{definition}

\begin{definition}[Decisional RLWE]
Given oracle access to i.i.d. samples, distinguish whether they come from $\mathsf{A}_{q,\chi}(s)$ for a secret $s$ drawn from a fixed distribution $D$ over $R^\vee$ (mod $q$) or from the uniform distribution on $R_q\times R_q^\vee$.
\end{definition}

\subsubsection{BCH and Hensel Lifting}
\label{subsec:prelim-bch-hensel}
Let $n=2^{m}-1$ and $\alpha\in\mathbb{F}_{2^{m}}$ be primitive. For designed distance $\delta\ge 2t+1$ and a consecutive index set $T=\{b,\ldots,b+\delta-2\}\ (\bmod n)$, the binary BCH code $\mathrm{BCH}(n,k,t)$ is the cyclic code generated by
$g_2(X)=\operatorname{lcm}\{M_i(X):i\in T\}$, where $M_i$ is the minimal polynomial of $\alpha^i$ over $\mathbb{F}_2$. Then $k=n-\deg g_2$ and $d_{\min}\ge \delta$~\cite{bose1960class}. A systematic encoder maps a message $u(X)$ to $c(X)=u(X)X^{n-k}+\bigl(u(X)X^{n-k}\bmod g_2(X)\bigr)$.
Algebraic decoding uses syndromes, Berlekamp--Massey (or EEA), and Chien search. For odd $N$, over $\mathbb{F}_2$ we have $X^{N}+1\equiv X^{N}-1$, so we may choose a binary BCH generator $g_2\mid (X^{N}+1)$ from a consecutive root set modulo $N$. Since $\gcd\!\bigl(g_2,(X^{N}+1)/g_2\bigr)=1$ in $\mathbb{F}_2[X]$, Hensel’s lemma lifts $g_2$ uniquely to a monic $g_k\in\mathbb{Z}_{2^{k}}[X]$ with $g_k\mid (X^{N}+1)$ and $g_k\equiv g_2\pmod{2}$.

\subsection{Non-leveled FHE Construction}
\label{sec:scheme}

In this subsection, we introduce a \emph{non-leveled} CKKS-based scheme that preserves the standard plaintext/ciphertext structure and programming model, while replacing leveled modulus switching with a \textsc{Refresh} and fixed-scale management. We retain the canonical embedding, RLWE-based encryption, and SIMD packing; the only structural change is to eliminate \texttt{Rescale} and instead invoke a periodic \textsc{Refresh}, a homomorphic re-encryption with Gaussian flooding, that maintains a single modulus and a fixed scale \(\Delta\). We first present the high-level workflow and then a concrete instantiation of this variant.

Our analysis shows that, without a modulus chain, the scheme controls noise growth under an explicit threshold, achieves quasi-linear complexity in the ring degree \(N\) (via NTT-based polynomial arithmetic), incurs only modest memory overhead, and supports homomorphic evaluation of general analytic functions.

\subsubsection{Framework Construction}
Similar to CKKS, our variant of CKKS's decryption follows the simple form \(\langle c, \mathsf{sk} \rangle = m + e\), where \(e\) is a small noise term. To ensure security, the encryption operation intentionally introduces a controllable small noise. During homomorphic computations, this error is monitored and refreshed when it exceeds the threshold to prevent error accumulation. While CKKS employs a rescaling mechanism to discard imprecise least significant bits (LSBs) and manage noise, our framework instead relies on an elementary bootstrapping to suppress noise growth.

Algorithm~\ref{alg:binary-ckks} is a concrete construction of our framework based on CKKS tailored for approximate computation over Gaussian integers with bounded-coefficient polynomial encoding and decoding. Correctness and some essential lemmas for error bounding are in the appendix.

\begin{algorithm}[!t]
\caption{Concrete Framework Construction}
\label{alg:binary-ckks}
\small
\begin{algorithmic}[1]

\STATE \textbf{Procedure} \textsc{KeyGen}$(1^{\lambda})$
\STATE $M \gets M(\lambda)$,\; $h \gets h(\lambda)$,\; $P \gets P(\lambda)$,\; $\sigma \gets \sigma(\lambda)$
\STATE $\tau \gets 2^{\kappa}\!\cdot\! B_{\max}$ for some $\kappa$
\STATE \textbf{Sample} $s\gets \mathrm{HWT}(h),\; a\gets R,\; e\gets \mathrm{DG}(\sigma^{2})$
\STATE $\mathsf{sk}\gets (1,s)$;\; $b \gets -\,a\cdot s + e$;\; $\mathsf{pk}\gets (b,a)$
\STATE \textbf{Relinearization key:}\; $a_0\gets R,\; e_0\gets \mathrm{DG}(\sigma^{2}),\;
       b_0\gets -\,a_0 s + e_0 + s^{2}$;\; $\mathsf{evk}\gets (b_0,a_0)$
\STATE \textbf{Boot key:}\; $\mathsf{bk}\gets \textsc{Enc}(\mathsf{pk},\,s)$
\STATE \textbf{return}\; $(\mathsf{sk},\mathsf{pk},\mathsf{evk},\mathsf{bk},\tau)$

\vspace{0.35em}
\STATE \textbf{Procedure} \textsc{Encode}$(z;\,\Delta)$
\REQUIRE $z=(z_j)_{j\in T}\in \mathbb{Z}[i]^{N/2}$
\STATE $c \gets \big\lfloor \Delta\;\pi^{-1}(z)\big\rceil_{\sigma(R)}$
\STATE \textbf{Apply inverse canonical embedding} to $c$ to obtain $m(X)$
\STATE \textbf{return}\; $m(X)$

\vspace{0.35em}
\STATE \textbf{Procedure} \textsc{Decode}$(m(X);\;\Delta)$
\STATE $z_j \gets \big\lfloor \Delta^{-1} m(\zeta_M^j)\big\rceil$ for $j\in T$
\STATE \textbf{return}\; $z=(z_j)_{j\in T}$

\vspace{0.35em}
\STATE \textbf{Procedure} \textsc{Enc}$(m(X))$
\STATE \textbf{Sample}\; $v\gets \mathrm{ZO}(0.5)$,\; $e_{0},e_{1}\gets \mathrm{DG}(\sigma^{2})$
\STATE \textbf{return}\; $v\cdot \mathsf{pk} + \bigl(m(X)+e_{0},\,e_{1}\bigr)$

\vspace{0.35em}
\STATE \textbf{Procedure} \textsc{Dec}$(c=(b,a))$
\STATE \textbf{return}\; $b + a\cdot s$

\vspace{0.35em}
\STATE \textbf{Procedure} \textsc{Add}$\big((c_{1},\Delta_1),(c_{2},\Delta_2)\big)$ \quad \emph{with } $\Delta_1\le \Delta_2$
\STATE $c_{\text{add}} \gets \textsc{MultConst}\!\left(\frac{\Delta_2}{\Delta_1},c_1\right) + c_2$;\; set $\Delta_1\!\to\!\Delta_2$
\STATE \textbf{return}\; $\textsc{MultConst}\!\left(\frac{\Delta}{\Delta_2},\,c_{\text{add}}\right)$

\vspace{0.35em}
\STATE \textbf{Procedure} \textsc{Mult}$\big(c_{1}=(b_{1},a_{1}),\,c_{2}=(b_{2},a_{2})\big)$
\STATE $(d_{0},d_{1},d_{2}) \gets \bigl(b_{1}b_{2},\; a_{1}b_{2}+a_{2}b_{1},\; a_{1}a_{2}\bigr)$
\STATE $c_{\text{mult}} \gets (d_{0},d_{1}) + d_{2}\cdot \mathsf{evk}$;\quad $\Delta_{\text{mult}}\!\to\!\Delta_1\Delta_2$
\STATE \textbf{return}\; $\textsc{MultConst}\!\left(\frac{\Delta}{\Delta_1\Delta_2},\,c_{\text{mult}}\right)$

\vspace{0.35em}
\STATE \textbf{Procedure} \textsc{Thresh}$(B_{\max},\,B_0)$
\STATE \textbf{return}\; $(B_0 > B_{\max})$

\vspace{0.35em}
\STATE \textbf{Procedure} \textsc{Refresh}$(c=(c_0,c_1),\,\mathsf{bk},\,\Delta,\,\tau)$
\STATE \emph{Homomorphic decrypt: obtain $\mathrm{Enc}(m{+}e)$}
\STATE $t \gets \textsc{MultConst}(c_1,\,\mathsf{bk})$ \hfill (equals $\mathrm{Enc}(c_1\!\cdot\! s)$)
\STATE $u \gets \textsc{AddConst}(t,\,c_0)$ \hfill ($u=\mathrm{Enc}(m{+}e)$)
\STATE \emph{Approx.\ rounding: choose $R$ with $R(x)\approx \mathrm{round}(x/\Delta)$}
\STATE $w \gets \textsc{EvalPoly}(u,\,R)$ \hfill ($w\approx \mathrm{Enc}(\mathrm{round}((m{+}e)/\Delta))$)
\STATE \emph{Re-encode to target scale}
\STATE $c_{\text{fresh}} \gets \textsc{MultConst}(\Delta,\,w)$
\STATE \emph{Re-randomize (circuit privacy)}
\STATE sample $v\!\leftarrow\!\mathrm{ZO}(0.5)$, $e_0,e_1\!\leftarrow\!\mathrm{DG}(\tau^2)$;\; $c_{\text{fresh}}\!\leftarrow\! c_{\text{fresh}} + v\cdot\mathsf{pk} + (e_0,e_1)$
\STATE \textbf{return}\; $c_{\text{fresh}}$

\end{algorithmic}
\end{algorithm}

Recall in standard CKKS~\cite[Section.~3.4~Lemma 3]{cheon2017homomorphic}, 
\[B_{mult}^{std}=
\nu_{1}B_{2}+\nu_{2}B_{1}+B_{1}B_{2}+P^{-1}q_{\ell}8\sigma N/\sqrt{3} + N+8/3\sqrt{h}.\]
The terms $\nu_1$ and $\nu_2$ represent rescaling factors applied to the plaintext values encoded in the two input ciphertexts, while $B_1$ and $B_2$ denote the corresponding upper bounds on their noise magnitudes. The parameter $P$ refers to the modulus employed during rescaling in standard CKKS, and $q_\ell$ is the modulus at level $\ell$ in the modulus chain. 
Given Lemma~\ref{multi}, we can compare the noise growth behavior of standard CKKS with our variant. Given $(c_1,B_1), (c_2,B_2)$, in our CKKS variant (eliminate $\Delta$ for a fair compare):
\begin{align*}
 B_{mult}^{var}=  \nu(B_2+ B_1)+B_1B_2+16\sqrt{3}\sigma^2 N.  
\end{align*}
To determine when the noise resulting from our CKKS variant's multiplication operation is smaller than that of the standard CKKS, we have the following Proposition~\ref{pro:bin_vs_std}.
\begin{proposition}
\label{pro:bin_vs_std}
Let $c_1$ and $c_2$ be \emph{fresh} ciphertexts in both schemes with identical $B_1 = B_2 = B_{\mathrm{enc}}$, $\nu_1 = \nu_2 = \Delta$, $B = \tfrac{\Delta}{2}$, and $h$ is the Hamming weight from key sampling. Denote the worst-case bound returned by the respective lemmas as
$B_{\mathrm{mult}}^{\mathrm{std}}$ and $B_{\mathrm{mult}}^{\mathrm{var}}$. We have the following:
\[
16\sqrt{3}\sigma^2 <\frac{8q_{\ell}}{\sqrt{3}P}+\frac{8\sqrt{h}}{\sqrt{3}\sqrt{N}} +1 \implies B_{\mathrm{mult}}^{\mathrm{var}}
    \;<\;
    B_{\mathrm{mult}}^{\mathrm{std}}.\] 
\end{proposition}
\noindent Proof in Appendix~\ref{proof_binvsstd}.\qed

This inequality holds trivially under the setting of CKKS, where $q_{\ell}$ is sufficiently large, proving the reduction of noise magnitude of the proposed non-level CKKS scheme. Lastly, let $c$ be an encryption of a message $m$ with noise at most $B$.
Then $\mathsf{Refresh}(\mathsf{pk},c)$ outputs a ciphertext $c'$ that is computationally indistinguishable from a fresh encryption of a rounded message
$m'$, with noise at most $B_r \;=\; \alpha B + \beta$, where $\alpha,\beta$ determined by the initial parameters as shown in Appendix~\ref{subsec:nonleveled-correctness} .

\subsubsection{Support for Analytic Functions}
We present algorithms for evaluating common circuit components in practical applications and analyze the associated error growth under the construction of our non-leveled variant. We begin with fundamental homomorphic operations, such as addition and multiplication by constants, monomials, and polynomials, which serve as building blocks for approximating analytic functions. For simplicity in analysis, we assume that the error introduced by multiplication and refresh operations is bounded by a constant \( B^* \). That is, if two ciphertexts \((c_1, B_1)\) and \((c_2, B_2)\) are multiplied and refreshed, the resulting ciphertext \((c_0, B_0)\) satisfies the relation: $B_0 \leq \min\{B^*,B_{mult}\}$. If $B_0$ reaches threshold $B^*$, we set $B_0$ as $B_{enc}$ with refresh operation. Analyzing basic constant operations, addition and multiplication by a scalar constant \( a \in R\), is shown as lemmas in the appendix, with extension to general polynomial evaluation further.

We describe Algorithm~\ref{poly}for homomorphic evaluation of power functions \( f(x) = x^d \), where \( d \) is a power-of-two integer, and analyze the error growth. At each squaring step, the ciphertext error approximately doubles, accumulates quadratic terms, and may trigger a refresh that resets noise to $B_{enc}$. For ciphertexts remaining below the threshold, the update rule is:
\[
B_j=\frac{2\nu B_{j-1}+B_{j-1}^2+16\sqrt{3}\sigma^2 N}{\Delta},
\]
where $B_j$ is the noise bound after squaring the original ciphertext with noise bound $B_{j-1}$. We can extend Algorithm 3 to polynomial evaluations.

\begin{algorithm}[!t]
\caption{Power polynomial $f(x)=x^{d}$ for $d=2^r$}
\label{poly}
\small
\begin{algorithmic}[1]
\STATE \textbf{Input:} ciphertext $c$, bound $B_c$, exponent $d=2^r$
\STATE \textbf{Output:} ciphertext $c_r$, bound $B_r$

\STATE $c_0 \gets c$
\STATE $B_0 \gets B_c$

\FOR{$j=1$ \TO $r$}
  \STATE $c_j \gets \textsc{Mult}(c_{j-1},\,c_{j-1})$
  \IF{$\dfrac{2\nu B_{j-1}+B_{j-1}^2+16\sqrt{3}\sigma^2 N}{\Delta} < B^{*}$}
    \STATE $B_j \gets \dfrac{2\nu B_{j-1}+B_{j-1}^2+16\sqrt{3}\sigma^2 N}{\Delta}$
  \ELSE
    \STATE $B_j \gets B_{\mathrm{enc}}$;\quad $c_j \gets \textsc{Refresh}(c_j)$
  \ENDIF
\ENDFOR

\RETURN $(c_r,\,B_r)$
\end{algorithmic}
\end{algorithm}

Let \( f(x) = \sum_{j=0}^\infty a_j x^j \) be a real analytic function on an interval containing the plaintext range. By definition, such an \(f\) admits a convergent power series expansion around \(0\). We truncate this series to degree \(D\) so that the tail error $\left| \sum_{j > D} a_j m^j \right| \leq \epsilon$, for all plaintext values \(m\) in the domain of interest. For example, if \(|m| \leq Q\), one can use a remainder bound from Taylor’s Theorem to choose \(D\) such that
\begin{align*}
|a_{D+1}| \cdot |m|^{D+1}/(D+1)! < \epsilon
\quad \text{or} \quad
\sum_{j>D} |a_j| Q^j \leq \epsilon.
\end{align*}
We evaluate the truncated polynomial \( P_D(x) = \sum_{j=0}^{D} a_j x^j \) on ciphertexts using an adaptive-refresh strategy. For each monomial \(a_j x^j\), the term \(x^j\) is computed via repeated squaring as in Algorithm~\ref{poly}. In practice, intermediate powers of \(x\) can be reused across terms. 

When \(\Delta\) is not set sufficiently large, each nontrivial multiplication rapidly approaches the threshold \(B^*\), triggering a \texttt{refresh} operation that resets the noise to the fresh encryption bound \(B_{\text{enc}}\), as shown in the proof of Lemma~6. We use this as the worst-case bound, though in practice, parameters can be chosen such that \texttt{refresh} is invoked less frequently, which requires a careful parametrization. The following theorem describes the evaluation of general analytic functions.
\begin{theorem}
\label{thm_analy}
Let function \( f(x) = \sum_{j=0}^\infty a_j x^j \) be analytic on an interval containing the plaintext domains, and suppose \((c_0, B_0)\) is a ciphertext encrypting \(m\) with initial noise \(B_0 = B_{\text{enc}}\). Fix a refresh threshold \(B^*\). Choose a truncation degree \(D\) depending on $\epsilon$ such that \(|\sum_{j>D} a_j m^j|<\epsilon\) for all \(m\) in the domain. Then the adaptive Algorithm~\ref{poly} outputs a ciphertext \((c_f, B_f)\) encrypting the approximate value \(P_D(m)\), such that the ciphertext noise satisfies
\[
B_f \approx \min\{ \sum_{j=0}^{D} |a_j| \cdot B_{\text{enc}} + \underbrace{\epsilon}_{\text{tail error}},\; B^* \}.
\]
\end{theorem}
\noindent Proof in Appendix~\ref{pf_analy}.\qed

\subsubsection{IND-CPA Security}
\label{indcpa-bckks}

We prove IND-CPA security for the encryption layer of the scheme. Let
\(D_s\) denote the secret-key distribution and \(D_v\) denote the
encryption-randomness distribution used for \(v\). We assume decisional
RLWE over \(R_q\) for secrets sampled from \(D_s\), and the standard
multi-sample decisional RLWE assumption for secrets sampled from \(D_v\).
The latter is the usual assumption needed for RLWE public-key encryption,
since a ciphertext contains RLWE samples whose ephemeral secret is the
encryption randomness \(v\).

\begin{theorem}
\label{thm:indcpa}
Assume decisional RLWE is hard over \(R_q\) for the secret distributions
\(D_s\) and \(D_v\). Then the encryption layer of the proposed
non-leveled CKKS construction is IND-CPA secure.
\end{theorem}
\noindent Proof in Appendix~\ref{pf_indcpa}.\qed

\paragraph{Auxiliary evaluation keys.}
The theorem above proves IND-CPA security of fresh ciphertexts under the
ordinary public key. If \(\mathsf{evk}\) and \(\mathsf{bk}\) are published
as part of the public evaluation material, then the full FHE public key
contains encryptions or RLWE encodings of functions of the secret key,
such as \(s\) and \(s^2\). Security of the full evaluated scheme therefore
requires the standard auxiliary-key/circular-security assumption used in
RLWE-based FHE schemes: the published evaluation and refresh keys are
computationally indistinguishable from uniformly random ring elements, even
when their embedded plaintexts are key-dependent functions. Under this
auxiliary-key assumption, the above IND-CPA proof extends to the public
evaluation-key setting.
%\paragraph{Multi‑client setting} If the secret key is additively shared,
%$s=\sum_{i=1}^{n}s_i$, each client publishes its own
%$\mathsf{rk}_i=Enc_{pk}(s_i)$.  
%During refresh, the evaluator homomorphically computes
%\(
%t = b+\sum_i a_i s_i
%\)
%using all $\mathsf{r}_i$, then re‑encrypts as above.
%Because every $s_i$ appears only under encryption
%and the flooding mask statistically hides $t$,
%any collusion of up to $n{-}1$ clients (or the evaluator)
%learns no additional information.
\subsubsection{Complexity and Memory Usage Analysis}
\label{spaceusage}
We analyze the asymptotic cost of the proposed non-leveled CKKS. Implementation uses an RNS modulus $q=\prod_{i=1}^{S} q_i$. Relinearization/key-switching employs gadget base $T\!\ge\!2$ with digit count $L=\lceil \log_T q\rceil$. When bootstrapping is enabled, the rounding polynomial has degree $d$ and $\rho$ slot rotations. With NTT, one degree-$N$ multiplication in $R_q$ costs $\mathcal{O}(S\,N\log N)$. Per-algorithm costs appear in Table~\ref{complexitybckks}; for fixed $S$ and $L$, runtime is quasi-linear in $N$ and dominated by $\widetilde{\mathcal{O}}((1{+}L)\,S\,N\log N)$ from multiplications.
Memory is linear in the number of stored ring polynomials: a degree-1 ciphertext holds two polys ($\approx 2N\Lambda$ bits); the evaluation key $\approx 2L\,N\Lambda$ bits; the boot key $\approx 2N\Lambda$ bits; rotation keys $\approx (2L)\,|\mathcal{G}|\,N\Lambda$ bits, where $|\mathcal{G}|$ is the number of Galois elements.

\begin{table}[!t]
\caption{Complexity Analysis of Non-Leveled CKKS Components (single fixed $q$)}
\label{complexitybckks}
\footnotesize
\setlength{\tabcolsep}{4pt}
\renewcommand{\arraystretch}{1.15}
\begin{tabular}{@{} l l p{0.50\columnwidth} @{}}
\toprule
\textbf{Procedure} & \textbf{Time (ring ops)} & \textbf{Randomness (fresh draws)} \\
\midrule
\textbf{KeyGen}
& $\widetilde O\!\big((1{+}L)\,S\,N\log N\big)$
& \makecell[l]{HWT($h$) for $s$; $U{\times}2$ ($a,a_0$); $DG{\times}2$ ($e,e_0$);\\
bk: Bern $+$ $DG{\times}2$.} \\
\addlinespace[0.2em]
\textbf{Ecd}
& $\widetilde O(N\log N)$
& none \\
\textbf{Dcd}
& same as \textbf{Ecd}
& none \\
\textbf{Enc}
& $\widetilde O(S\,N\log N)$
& Bern ($v$); $DG{\times}2$ ($e_0,e_1$) \\
\textbf{Dec}
& $O(SN)$
& none \\
\textbf{Add}
& $O(SN)$
& none \\
\textbf{Mult} (incl.\ rlin)
& $\widetilde O\!\big((1{+}L)\,S\,N\log N\big)$
& none (uses $\mathsf{evk}$) \\
\textbf{Refresh} (bootstrap)
& $\widetilde O\!\big((\sqrt d{+}\rho)\,S\,N\log N\big)$
& optional flooding: Bern $+$ $DG{\times}2$ \\
\midrule
\multicolumn{3}{@{}p{\columnwidth}@{}}{\footnotesize
\textbf{Params:} $S$ = \#RNS limbs ($q=\prod q_i$), $L=\lceil\log_T q\rceil$ (gadget digits), $d$ = degree of rounding polynomial, $\rho$ = \#slot rotations.
Each ring polynomial stores $\approx N\Lambda$ bits with $\Lambda=\log_2 q$. The boot key (bk) is a single ciphertext $Enc(s)$.
}\\
\bottomrule
\end{tabular}
\end{table}

\subsection{Fault-Tolerance via Ring--BCH}
\label{sec:ring-bch}
While CKKS is attractive for privacy-preserving ML due to its support for approximate arithmetic and SIMD packing, practical deployments demand robust decryption in the presence of noise from both homomorphic evaluation and real-world transport. A single RNS-limb fault can compromise confidentiality, and in production ML pipelines, such as streaming/RPC, federated aggregation, batched inference, transient bit flips during data transit can derail decoding and force costly retries or re-transmissions. To address these issues, we introduce an algebraic error-correcting layer that stays inside the plaintext ring and composes cleanly with non-leveled CKKS.

\begin{figure}[!t]
\centering
\includegraphics[width=1.1\linewidth, trim=1cm 1cm 1cm 1cm, clip]{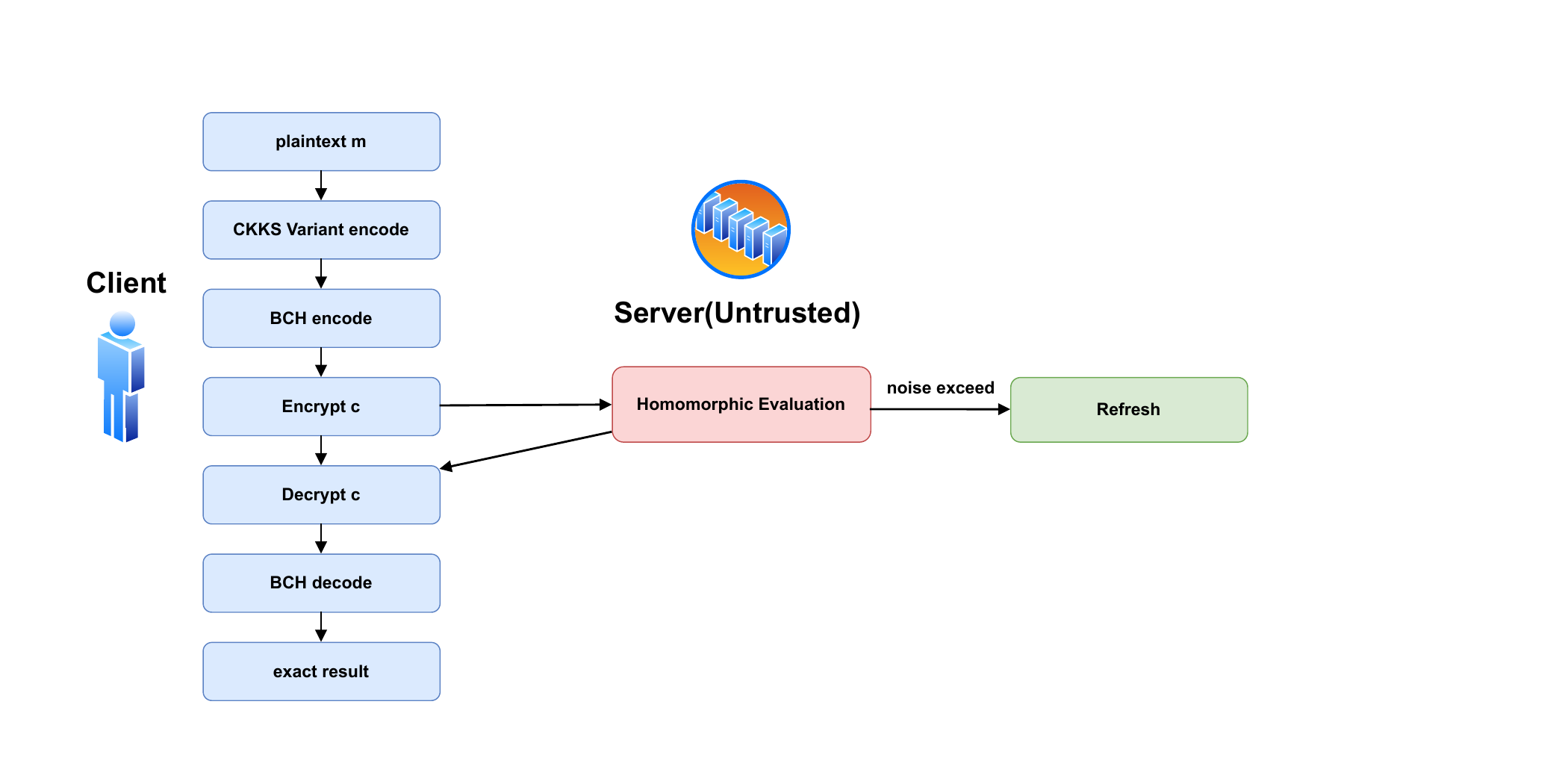}
\caption{Ring--BCH pre-/post-processing entirely in $R=\mathbb{Z}_{q}[X]/(X^N{+}1)$.}
\label{fig:ring-bch-wf}
\end{figure}

Concretely, inspired by~\cite{moreira2006essentials, aly2007quantum,forney2003decoding}, we instantiate a Ring–BCH code as an ideal of $R=\mathbb{Z}_{2^k}[X]/(X^N{+}1)$, ensuring closure under the ring operations that underlie linear homomorphic evaluation. We further employ a code-preserving automorphism interleaver to disperse low-significance error clusters without leaving the code. This reliability layer supports privacy-preserving linear algebra and integrates into end-to-end ML workflows; it can be extended to broader HE circuits by exploiting the algebraic structure of $R$. Figure~\ref{fig:ring-bch-wf} illustrates the workflow: messages are segmented into $R$-polynomial blocks, encoded by a Ring–BCH ideal, interleaved via a ring automorphism, encrypted, and finally decoded after decryption.

\subsubsection{Construction Overview}
\label{subsec:ring-bch}
We work in the $2$-adic setting with odd $N$ such that $N\mid(2^m{-}1)$, so $X^N{+}1$ splits over $\mathbb{F}_{2^m}$. Using the standard binary BCH construction with designed distance $\delta$ and consecutive locators, we obtain a generator $g_2\in\mathbb{F}_2[X]$ dividing $X^N{+}1$. By Hensel lifting there is a unique monic $g_k\in\mathbb{Z}_{2^k}[X]$ with $g_k\bmod 2=g_2$ and $g_k\mid(X^N{+}1)$. We define the Ring–BCH code as the principal ideal $(g_k)$ in $R=\mathbb{Z}_{2^k}[X]/(X^N{+}1)$ and use systematic encoding with payload dimension $K=N-\deg g_k$.

This choice is operationally convenient for PPML pipelines. Because $(g_k)$ is an ideal of $R$, it is closed under ring addition and multiplication, so linear homomorphic evaluation preserves code membership and does not interfere with encrypted linear algebra. Reducing modulo $2$ recovers the underlying binary BCH code with designed distance at least $\delta$, and standard Gray-map arguments transfer corresponding Lee/Hamming distance lower bounds over $\mathbb{Z}_{2^k}$, providing deterministic correction capability. Finally, ring automorphisms act as coordinate permutations that preserve the ideal, dispersing burst errors before decoding without leaving the code~\cite{deng2022random}. In practice, we pack $K$ slots per block, evaluate homomorphically as usual, and decode at the edge with low overhead while retaining the CKKS programming model.

\subsubsection{Systematic encoding and decoding.}
For a message $u(X)$ with $\deg u<K$, define the systematic codeword
\[
c(X) \;=\; u(X)\,X^{N-K} \;+\; \big(u(X)\,X^{N-K}\bmod g_k(X)\big)\ \in\ \mathcal{C}.
\]
Decoding proceeds by a mod-2 pass (syndromes, BM/EEA, Chien, Forney) followed by Hensel lifting of the error values to $2^k$ and inversion of the systematic map. All steps run \emph{in $R$}. A concrete encoder/decoder appears in Algorithm~\ref{alg:ring-bch} (appendix A).

\subsubsection{Interleaving}
\begin{figure}[!t]
\includegraphics[width=1.1\linewidth, trim=3cm 15cm 0cm 0cm, clip]{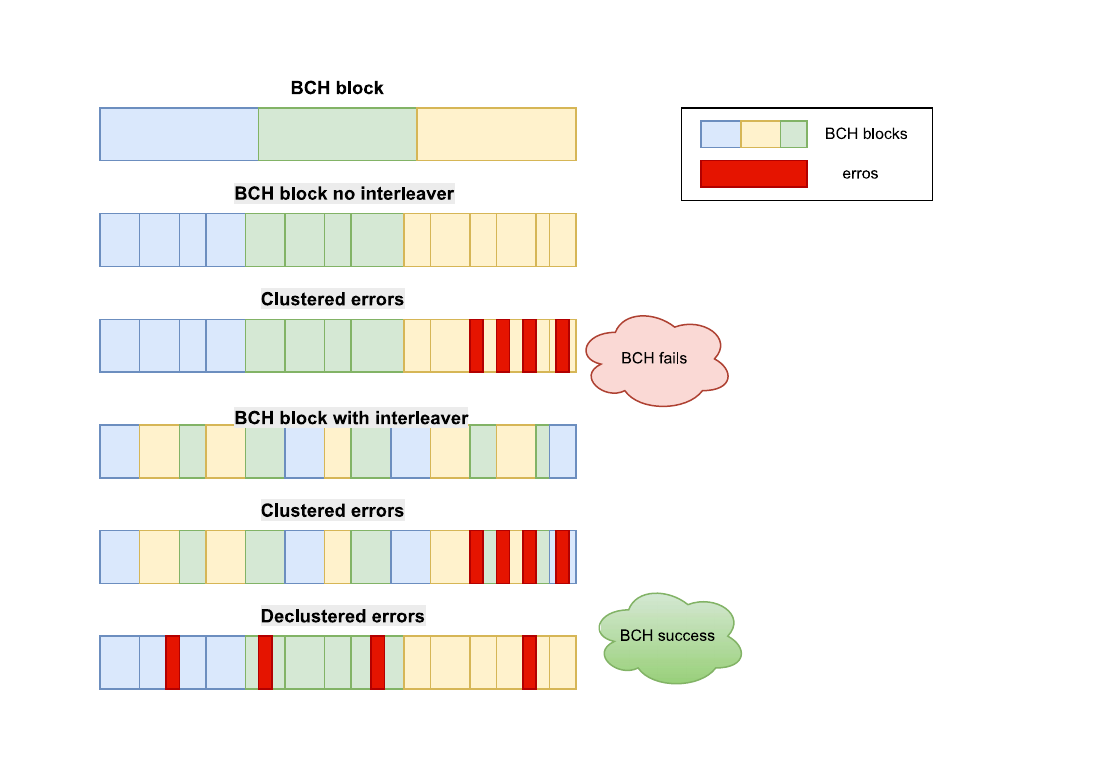}
\caption{Interleaving de-clusters errors: yellow/blue/green are BCH blocks; red stripes are localized errors.}
\label{fig:LSB}
\end{figure}

CKKS error tends to concentrate in low significance within \emph{coefficients}. To de-cluster these errors across code positions, we use a ring automorphism as an algebraic interleaver. Figure~\ref{fig:LSB} illustrates the essence of this automorphism. For any odd $s$ with $\gcd(s,N)=1$, the map $\sigma_s:R\to R$, $X\mapsto X^s$, is a ring automorphism since
\(
(X^s)^N=(X^N)^s=(-1)^s=-1 \text{ in }R.
\)
Because $\mathcal{C}=(g_k)$ is an ideal, $\sigma_s(\mathcal{C})=\mathcal{C}$. We use $\sigma_s$ (and optionally cyclic shifts $X^t$) as \emph{code-preserving} interleavers that permute coefficient indices $i\mapsto s\,i\bmod N$ while keeping closure under ring addition/multiplication. At the sender, apply $\sigma_s$ to each codeword before encryption; at the receiver, apply $\sigma_{s^{-1}}$ prior to decoding.

\subsubsection{Payload Layout in $R$}
Let $M$ be the payload length in bits. Fix $\mathrm{BCH}(n,k_{\mathsf{BCH}},t)=(127,101,3)$ and select $n=127$ (or embed a degree-$127$ factor in a larger odd $N$ and restrict to that component). Segment the payload into $h \;:=\;\lceil \frac{M}{k_{\mathsf{BCH}}}\rceil$ blocks $u^{(i)}\in\{0,1\}^{k_{\mathsf{BCH}}}$, map each block to a polynomial $u^{(i)}(X)$ over $\mathbb{Z}_{2^k}$ by placing the bits in the LSBs of the $K$ message positions, encode $c^{(i)}=\textsc{RingBCH\_Encode}(u^{(i)})\in\mathcal{C}\subset R$, and optionally interleave via $c^{(i)}\leftarrow\sigma_s(c^{(i)})$. Each $c^{(i)}$ is then encrypted as a plaintext in $R$. On the way back, decrypt, apply $\sigma_{s^{-1}}$, decode with \textsc{RingBCH\_Decode}, and concatenate the recovered $u^{(i)}$’s. Algorithm~\ref{alg:seg-enc} in the appendix shows this design concretely.

\subsubsection{IND-CPA Security}
\label{subsec:ring-bch-indcpa}
Let $\mathcal{S}$ be the base scheme over $R$ and $\mathcal{S}^\star$ the scheme that composes Ring--BCH pre-encoding before encryption and Ring--BCH post-decoding after decryption. The Ring--BCH layer is public, deterministic, and invertible on codewords; therefore, it is a benign pre-/post-processing.

\begin{theorem}
\label{thm_indcpa_bch}
Let \(\mathcal S\) be the base encryption scheme and let
\(\mathcal S^\star\) be the scheme obtained by applying the public
deterministic Ring--BCH encoder before encryption and the corresponding
decoder after decryption. If \(\mathcal S\) is IND-CPA secure, then
\(\mathcal S^\star\) is IND-CPA secure. More precisely, for every PPT
adversary \(\mathcal A\) against \(\mathcal S^\star\), there exists a PPT
adversary \(\mathcal B\) against \(\mathcal S\) such that
\[
   Adv_{\mathcal S^\star}^{\mathsf{ind\mbox{-}cpa}}(\mathcal A)
  \le
   Adv_{\mathcal S}^{\mathsf{ind\mbox{-}cpa}}(\mathcal B).
\]
\end{theorem}
\noindent Proof sketch in Appendix~\ref{pf_indcpa_bch}.

\section{Evaluation}
\label{sec:evaluation}

We evaluate our framework from a system perspective, focusing on its suitability for web-scale and federated machine learning deployments. Specifically, we ask:

\begin{itemize}
   \item \textbf{Q1: Latency and throughput.} What is the measured overhead of
encrypted low-degree activation evaluation, and how much additional latency is
introduced by one invocation of our experimental Refresh skeleton?
   \item \textbf{Q2: Robustness.} How effectively does the Ring--BCH layer correct
bursty in-transit coefficient faults modeled by a Gilbert--Elliott channel?
    \item \textbf{Q3: Accuracy preservation.} Does our approach maintain model accuracy across common privacy-preserving ML tasks under realistic noise and error conditions?
\end{itemize}

\subsection{Experimental Setup}
We implemented our system on top of \textsc{HElib}~\cite{halevi2014algorithms} with custom modules for non-leveled CKKS and Ring–BCH encoding/decoding. All benchmarks were run on a Linux host with an AMD Ryzen 7 (8 cores/16 threads) CPU and 32 GB RAM, running Ubuntu 22.04. To emulate federated learning, we deployed client instances that perform local encryption and transmission, while the server aggregates and evaluates ciphertexts. We locally implemented the proposed framework by extending the HElib library with custom modules while making minimal changes to the library’s core. Specifically, we introduced two new source files $\texttt{simple\_nonleveled\_ckks.h}$ and $\texttt{simple\_nonleveled\_ckks.cpp}$, which define a class to support non-leveled CKKS operations. These new classes encapsulate the entire encryption-scheme workflow, including key generation, encoding, encryption, homomorphic evaluation, and decryption, all tailored for single-level data. We modified a few existing HElib utility headers (e.g., $\texttt{binio.h}$ and $\texttt{io.h}$) to accommodate our custom data types, while most of the added functionality resides in the new files within the $\texttt{benchmarks/}$ directory, keeping the core codebase intact. 

Overall, we added roughly
$1,100$ lines of code across two new files,
\texttt{simple\_\allowbreak nonleveled\_\allowbreak ckks.\allowbreak h} and
\texttt{simple\_\allowbreak nonleveled\_\allowbreak ckks.\allowbreak cpp}, and touched fewer than~40 lines in three existing \textsc{HElib} utility
headers (\texttt{binio.\allowbreak h}, \texttt{io.\allowbreak h}, and
\texttt{NumbTh.\allowbreak h}). Because virtually all new logic resides in
the two new files under \texttt{benchmarks/}, the core \textsc{HElib} codebase remains intact, making the modification easy to audit, maintain,
and port. Table~\ref{tab:apps-1col} lists the workload categories that motivate the
design. In the quantitative results below, we focus on encrypted low-degree
activation evaluation, one experimental Refresh skeleton invocation, and
transport-fault robustness of the Ring--BCH layer.

% ---------- Table 2: Datasets ----------
\begin{table}[t]
\footnotesize
\centering
\caption{Datasets for Application}
\label{tab:dataset}
\begin{tabular}{ll}
\toprule
Workload & Datasets \\
\midrule
Encrypted inference & MNIST (60k/10k), CIFAR-10 (50k/10k) \\
Federated aggregation & MNIST/CIFAR-10 with Dirichlet non-IID ($\alpha\!\in\!\{0.1,0.3\}$)\\
Streaming analytics & UCI Household Power (minute-level);\\
\bottomrule
\end{tabular}
\end{table}

% ---------- Table 3: Applications & Metrics ----------
\begin{table}[t]
\footnotesize
\caption{Models/operations and primary metrics (tested at $N\!\in\!\{1024,2048,4096,8192\}$).}
\label{tab:apps-1col}
\begin{tabular}{@{} p{0.26\linewidth} p{0.34\linewidth} p{0.34\linewidth} @{}}
\toprule
\textbf{Workload} & \textbf{Model / Operation} & \textbf{Primary Metrics} \\
\midrule
Encrypted inference & LogReg; 2-layer MLP  & Top-1; end-to-end latency; p99; throughput \\
Federated aggregation & Encrypted mean of client updates per round & Agg.\ latency vs.\ clients; p99 tail; bytes; rel.\ $\ell_2$ error \\
Streaming analytics & Sliding-window mean  & memory vs.\ $W$; failure vs.\ $p$; $\ell_\infty$ deviation \\
\bottomrule
\end{tabular}
\end{table}

\subsection{End-to-End Latency and Throughput}

We remeasured the encrypted activation experiment using HElib CKKS. Each
baseline run encrypts a packed input vector, evaluates the cubic activation
\[
f(x)=0.125x^3+0.25x^2+0.5x+0.125,
\]
decrypts the result, and compares against a plaintext reference. BCH transport
coding is excluded from this homomorphic-evaluation timing.

We use this degree-3 polynomial because CKKS supports additions and
multiplications directly, but not non-polynomial activations such as ReLU or
sigmoid. A cubic activation is a standard low-depth proxy: it is nonlinear,
uses only two ciphertext--ciphertext multiplications, and gives a controlled
benchmark for encrypted activation evaluation without requiring bootstrapping.
Thus, the polynomial is used to measure the cost of a realistic low-degree
encrypted nonlinear layer, not to claim it is the best approximation for every
ML model.

For power-of-two HElib CKKS parameters, HElib uses the cyclotomic index $m$.
The paper reports ring degree $N$. In these experiments we use the mapping
$m=2N$. Thus $N\in\{1024,2048,4096,8192\}$ corresponds to
$m\in\{2048,4096,8192,16384\}$, and the number of CKKS slots is $N/2$.

\begin{table}[t]
\centering
\caption{Measured CKKS encrypted cubic activation latency. BCH is excluded.}
\label{tab:sec42_ckks_activation_latency}

\setlength{\tabcolsep}{3.2pt}
\renewcommand{\arraystretch}{1.05}
\begin{tabular*}{\linewidth}{@{\extracolsep{\fill}}rrrrrrr@{}}
\toprule
$N$ & $m$ & Slots & Enc. & Eval. & Dec. & Total \\
    &     &       & ms & ms & ms & ms \\
\midrule
1024 & 2048  & 512  & 0.477 & 9.503  & 10.739 & 20.719  \\
2048 & 4096  & 1024 & 0.978 & 17.551 & 21.550 & 40.079  \\
4096 & 8192  & 2048 & 1.860 & 36.350 & 42.832 & 81.041  \\
8192 & 16384 & 4096 & 4.002 & 73.056 & 88.722 & 165.779 \\
\bottomrule
\end{tabular*}
\end{table}

We also benchmarked the proposed public homomorphic decrypt-and-reencrypt
Refresh construction. The benchmark precomputes encrypted secret-key material
$\operatorname{Enc}(s)$ and, for rerandomization, a precomputed encryption of
zero. The online Refresh skeleton evaluates
\[
u = c_0 + c_1\cdot \operatorname{Enc}(s),
\]
then applies a degree-3 polynomial surrogate for the approximate rounding /
EvalPoly step, rescales back to the target scale, and adds the precomputed
encrypted zero.\footnote{This is not yet a complete CKKS bootstrapping
implementation, because the polynomial surrogate is not a validated CKKS
EvalMod or approximate modular-reduction circuit.}

The degree-3 polynomial in the Refresh skeleton is used only as a low-depth
prototype for the approximate rounding step. It keeps the skeleton cheap and
comparable across parameters, while still exercising the encrypted polynomial
evaluation path. It should not be interpreted as a complete EvalMod circuit or
as a correctness proof for CKKS bootstrapping. Plain decrypt-and-reencrypt and
the public homomorphic Refresh skeleton are measured separately.

\begin{table}[t]
\centering
\caption{Public Refresh skeleton timing. The skeleton uses precomputed
$\operatorname{Enc}(s)$ and precomputed encrypted zero, and uses a degree-3
surrogate rather than a complete EvalMod circuit.}
\label{tab:sec42_refresh_audit}
\setlength{\tabcolsep}{3.4pt}
\renewcommand{\arraystretch}{1.05}
\begin{tabular*}{\linewidth}{@{\extracolsep{\fill}}rrrr@{}}
\toprule
$N$ & $m$ & Hom. dec core & Public Refresh skeleton \\
    &     & ms & ms \\
\midrule
1024 & 2048  & 0.046 & 7.178  \\
2048 & 4096  & 0.089 & 16.401 \\
4096 & 8192  & 0.166 & 31.753 \\
8192 & 16384 & 0.335 & 61.300 \\
\bottomrule
\end{tabular*}
\end{table}

Including one full Refresh skeleton invocation after the encrypted activation
gives the following experimental end-to-end timings.

\begin{table}[t]
\centering
\caption{Encrypted activation plus one experimental Refresh skeleton.}
\label{tab:sec42_activation_plus_refresh_skeleton}
\setlength{\tabcolsep}{3.4pt}
\renewcommand{\arraystretch}{1.05}
\begin{tabular*}{\linewidth}{@{\extracolsep{\fill}}rrrrrr@{}}
\toprule
$N$ & $m$ & CKKS base & Refresh skel. & Total & Thru./s \\
    &     & ms &  ms & ms & \\
\midrule
1024 & 2048  & 20.719  & 7.178  & 27.897  & 35.846 \\
2048 & 4096  & 40.079  & 16.401 & 56.481  & 17.705 \\
4096 & 8192  & 81.041  & 31.753 & 112.794 & 8.866  \\
8192 & 16384 & 165.779 & 61.300 & 227.079 & 4.404  \\
\bottomrule
\end{tabular*}
\end{table}

At $N=4096$, the linear homomorphic decryption core costs only $0.166$ ms,
but the fuller Refresh skeleton costs $31.753$ ms. This shows that the core
linear expression $c_0+c_1\cdot\operatorname{Enc}(s)$ is inexpensive; most of
the Refresh skeleton cost comes from the polynomial surrogate evaluation and
scale-restoration path.

\subsection{Robustness to Network Faults}
\label{sec:robust-bursty}
We evaluate Q2 under correlated transmission errors using a Gilbert--Elliott channel. In the Good state, each coefficient flips with base rate $p\!\in[10^{-9},10^{-5}]$; in the Bad state (packet/word corruption) the flip rate is $p_{\mathrm{bad}}{=}2{\times}10^{-2}$ with average burst length $L_{\mathrm{bad}}{=}64$ coefficients and average Good-run length $L_{\mathrm{good}}{=}800$. For the evaluated transport layer, we use BCH$(127,101,3)$ as an outer
coefficient-block code. At CKKS ring degree $N$, we protect the coefficient
payload using \(h=\lceil N/101\rceil\) BCH blocks, e.g., \(h=41\) at
\(N=4096\). This evaluated transport code is block-based; it should be
distinguished from the ideal-theoretic Ring--BCH construction described above,
which requires compatible odd code lengths. A trial succeeds iff every block has $\le t{=}3$ symbol errors. We compare \emph{No BCH}, \emph{BCH (no interleaver)}, and \emph{BCH+Interleave} (automorphism-based stride interleaver). Each point averages 600 Monte Carlo trials.

\begin{figure}[t]
  \centering
  \begin{minipage}{0.49\linewidth}
    \includegraphics[width=\linewidth, trim=4.5cm 12cm 2cm 4cm, clip]{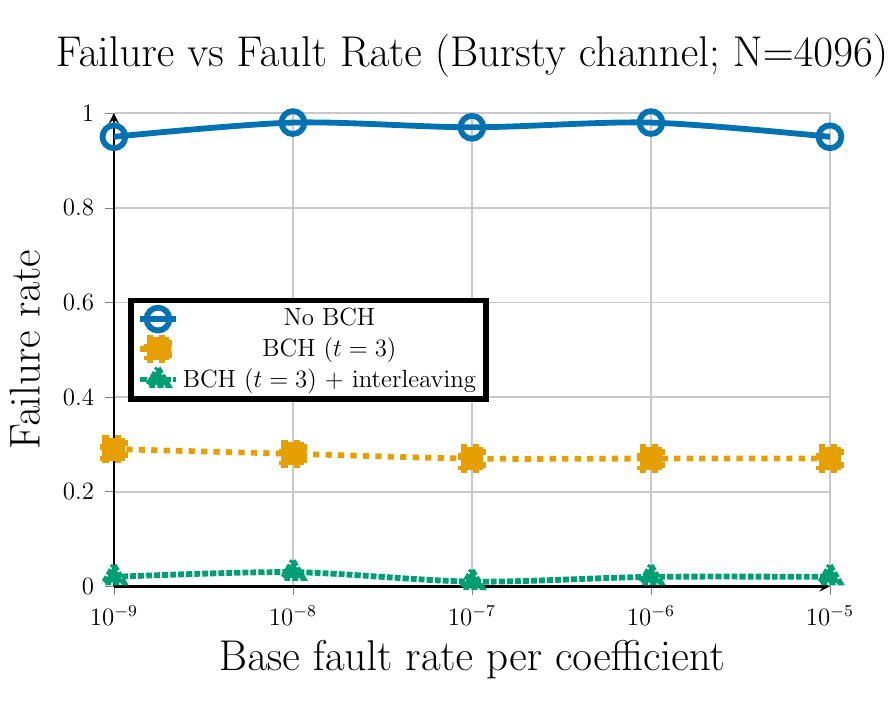}
    \vspace{0.2em}
    \centerline{\footnotesize (a) Failure vs.\ fault rate }
  \end{minipage}\hfill
  \begin{minipage}{0.49\linewidth}
    \includegraphics[width=\linewidth, trim=4.5cm 12cm 2.0cm 4cm, clip]{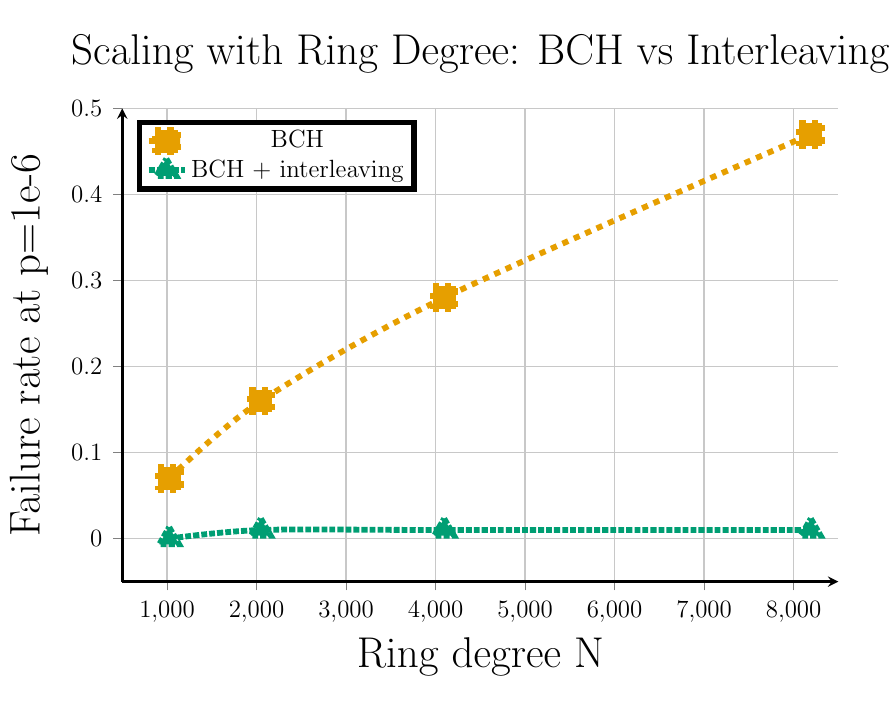}
    \vspace{0.2em}
    \centerline{\footnotesize (b) Failure at ($p{\approx}1.6{\times}10^{-6}$)}
  \end{minipage}\hfill

  \vspace{-0.4em}
  \caption{Robustness and cost under bursty faults. (a) Interleaving keeps failure $<0.5\%$. (b) Failure vs.\ $N$ without/with interleaving.}
  \label{fig:robustness-all}
\end{figure}

\paragraph{N=4096 across $p$.}
Figure~\ref{fig:robustness-all}(a) shows that \emph{No BCH} fails in $95.7$–$98.2\%$ of trials across $p\!\in[10^{-9},10^{-5}]$. BCH without interleaver improves to $25.3$–$30.2\%$ failure, while \emph{BCH+Interleave} is stable at $0.17$–$0.50\%$.

\paragraph{Scaling with $N$.}
At a representative base rate \(p{\approx}1.6{\times}10^{-6}\), BCH without interleaver failure grows from \(\mathbf{6.7}\%\) at \(N{=}1024\) to \(\mathbf{16.3}\%\) at \(2048\), \(\mathbf{29.5}\%\) at \(4096\), and \(\mathbf{47.8}\%\) at \(8192\); in contrast, \emph{BCH+Interleave} stays near \(0.17\%\)–\(0.33\%\) across all \(N\) (Fig.~\ref{fig:robustness-all}(b)). With 600 trials, 95\% Clopper--Pearson intervals are shown in Table~\ref{tab:fail-ci}. These results quantify the intuition that with more BCH blocks, the probability that some block exceeds the $t$-budget rises quickly unless bursts are dispersed.

% (acmart already loads booktabs)
\begin{table}[t]
\centering
\caption{Failure at $p\!\approx\!1.6\times10^{-6}$ (95\% CIs in parentheses; 600 trials).}
\label{tab:fail-ci}
\setlength{\tabcolsep}{3pt}
\renewcommand{\arraystretch}{1.05}
\begin{tabular*}{\linewidth}{@{\extracolsep{\fill}} lcc @{}}
\toprule
\textbf{$N$} & \textbf{BCH (\%)} & \textbf{BCH+Interleave (\%)} \\
\midrule
1024 & 6.7 (4.8–9.0)   & 0.17 (0.00–0.93) \\
2048 & 16.3 (13.5–19.5) & 0.33 (0.04–1.20) \\
4096 & 29.5 (25.9–33.3) & 0.17 (0.00–0.93) \\
8192 & 47.8 (43.8–51.9) & 0.17 (0.00–0.93) \\
\bottomrule
\end{tabular*}
\vspace{-0.6em}
\end{table}

\paragraph{Overhead vs.\ $N$.}
At $N{=}4096$ the total is \(4{,}160~\mu\text{s}\), decomposed as encode \(1{,}690~\mu\text{s}\), permute \(585~\mu\text{s}\), inverse \(552~\mu\text{s}\), and decode \(1{,}332~\mu\text{s}\). This adds only a few milliseconds to multiply-dominated inference pipelines ( \S\ref{sec:evaluation}) while yielding two orders of magnitude lower failure under bursts. At $N{=}8192$, the total overhead is $\approx 8.3 ms$, which is very economical.

\paragraph{Takeaways.}
(i) Without protection, virtually any burst derails decryption. (ii) BCH alone helps but degrades with ring dimension because the chance that one block exceeds $t$ grows with $h{=}\lceil N/101\rceil$. (iii) Automorphism-based interleaving keeps per-block errors within $t$, making failure almost \emph{flat in $N$}, at sub-percent levels under bursty faults, with millisecond-level overhead.

\subsection{Accuracy Preservation in ML Tasks}
\label{sec:acc}
We quantify the effect of Ring--BCH transport failures on modeled inference
accuracy. Following prior CKKS-style evaluations, we consider two representative
settings: a binary classification task and a 10-class task. Let
\(A_{\mathrm{plain}}\) denote plaintext test accuracy and \(A_{\mathrm{rand}}\)
the accuracy of a random guesser (\(0.5\) for binary and \(0.1\) for 10-class).
In this experiment, we isolate the effect of transport faults on downstream
accuracy. Since the Refresh skeleton is not yet a validated CKKS bootstrapping
implementation, we do not use this experiment to quantify Refresh approximation
error. Instead, we model the effect of decode failure: if a trial fails, we
conservatively assign a random prediction. Thus,
\[
A_{\mathrm{exp}}=(1-f)A_{\mathrm{plain}}+fA_{\mathrm{rand}},
\]
where \(f\) is the measured failure rate from \S\ref{sec:robust-bursty}.

\begin{figure}[t]
  \centering
  \begin{minipage}{0.49\linewidth}
    \includegraphics[width=\linewidth, trim=4.5cm 12cm 2.5cm 4cm, clip]{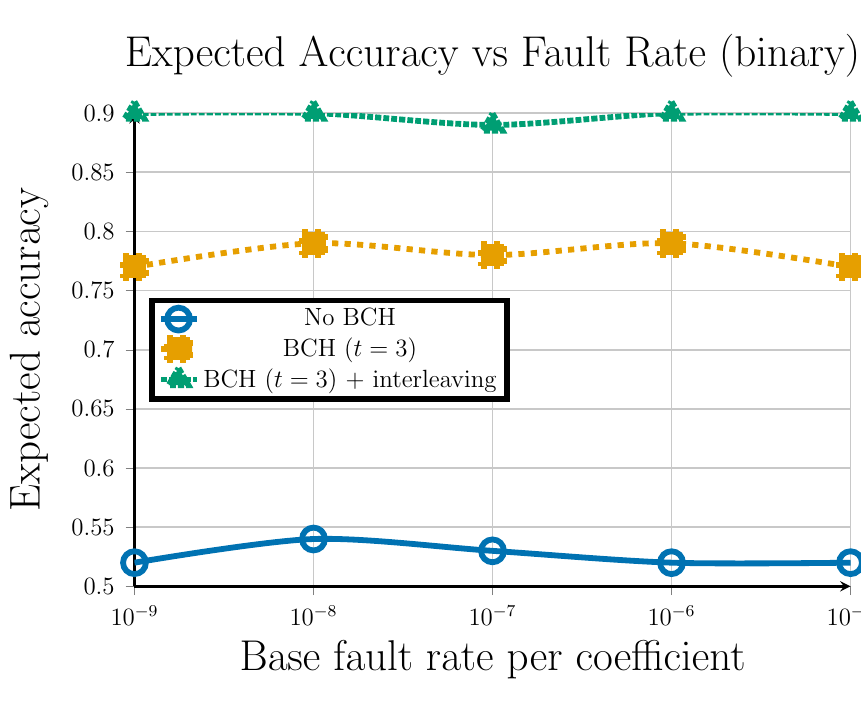}
    \vspace{-0.6em}\caption*{(a) Binary, $N{=}4096$}
  \end{minipage}\hfill
  \begin{minipage}{0.49\linewidth}
    \includegraphics[width=\linewidth, trim=4.5cm 12cm 2.5cm 4cm, clip]{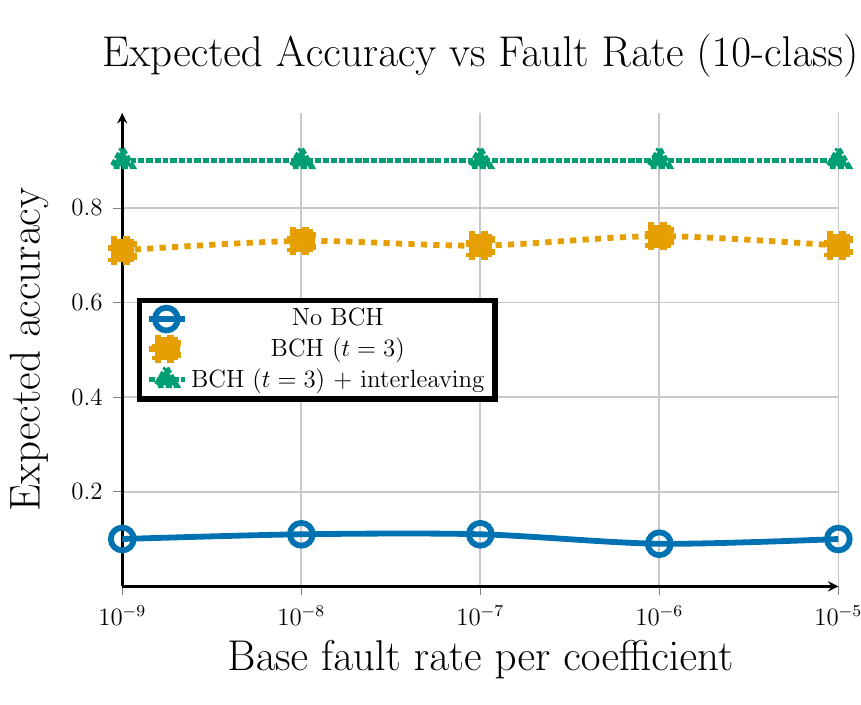}
    \vspace{-0.6em}\caption*{(b) 10-class, $N{=}4096$}
  \end{minipage}
  \vspace{-0.4em}
  \caption{Expected accuracy vs.\ base fault rate $p$ using failure rates from Fig.~\ref{fig:robustness-all}(a) and the model $A_{\mathrm{exp}}=(1{-}f)A_{\mathrm{plain}}+fA_{\mathrm{rand}}$. We set $A_{\mathrm{plain}}{=}0.90$ (binary) and $0.98$ (10- class) for illustration.}
  \label{fig:acc-vs-p}
\end{figure}

Using the bursty channel from \S\ref{sec:robust-bursty}, Fig.~\ref{fig:acc-vs-p} shows that BCH with Interleave tracks the plaintext baseline: the expected accuracy remains ${\ge}0.895$ (binary) and ${\ge}0.975$ (10- class) across $p\!\in[10^{-9},10^{-5}]$. In contrast, without interleaving, clustered errors reduce expected accuracy to $0.78$--$0.80$ (binary) and $0.72$--$0.76$ (10- class), reflecting the measured $25$--$30\%$ failure in Fig.~\ref{fig:robustness-all}(a). The \emph{No BCH} baseline collapses to the random guesser whenever bursts occur, yielding ${\sim}0.51$ (binary) and ${\sim}0.12$ (10- class).

\begin{figure}[t]
  \centering
  \begin{minipage}{0.49\linewidth}
    \includegraphics[width=\linewidth, trim=4.5cm 12cm 2.5cm 4cm, clip]{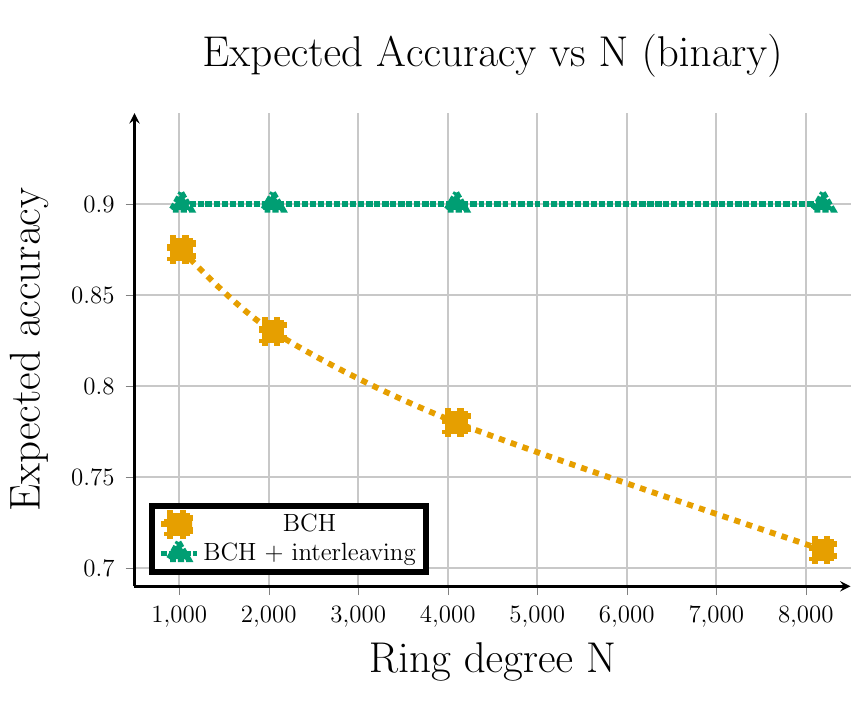}
    \vspace{-0.6em}\caption*{(a) Binary, $p{\approx}1.6{\times}10^{-6}$}
  \end{minipage}\hfill
  \begin{minipage}{0.49\linewidth}
    \includegraphics[width=\linewidth, trim=4.5cm 12cm 2.5cm 4cm, clip]{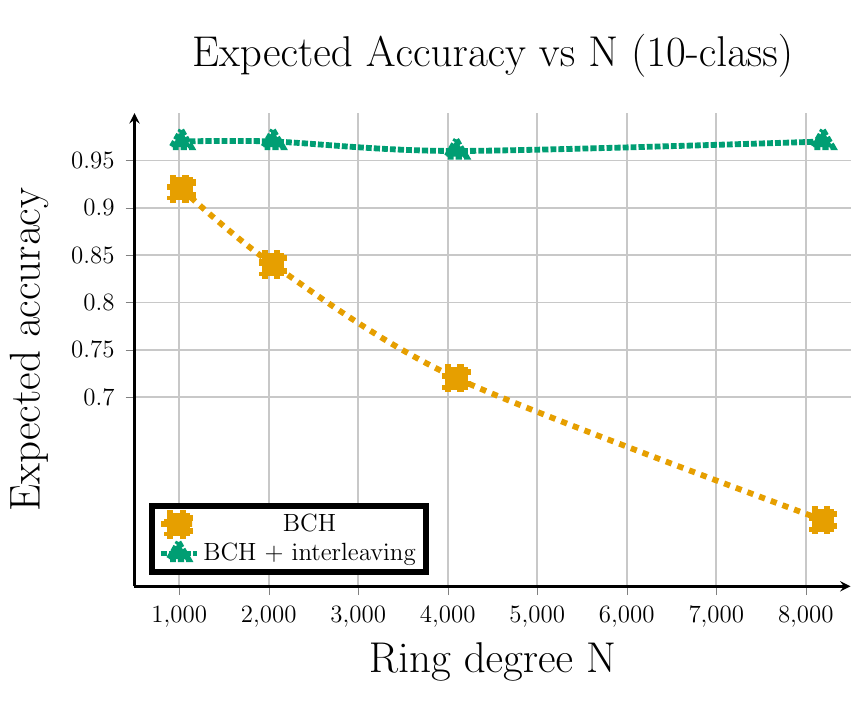}
    \vspace{-0.6em}\caption*{(b) 10- class, $p{\approx}1.6{\times}10^{-6}$}
  \end{minipage}
  \vspace{-0.4em}
  \caption{Expected accuracy vs.\ ring degree $N$ at a representative base rate ($p{\approx}1.6{\times}10^{-6}$).
  BCH (no interleaver) degrades with $N$ because the number of blocks $h{=}\lceil N/101\rceil$ grows, increasing the chance that some block exceeds the $t{=}3$ budget; interleaving keeps accuracy near the plaintext baseline.}
  \label{fig:acc-vs-N}
\end{figure}

\noindent\textbf{Scaling with $N$.}
At $p{\approx}1.6{\times}10^{-6}$, expected accuracy for BCH without
interleaver declines as $N$ grows: binary drops from
\[
0.873{\rightarrow}0.835{\rightarrow}0.782{\rightarrow}0.709
\]
at $N{=}1024,2048,4096,8192$; 10-class drops from
\[
0.921{\rightarrow}0.837{\rightarrow}0.720{\rightarrow}0.559.
\]
In contrast, BCH$+$Interleave remains flat within ${<}0.5$ percentage points
of plaintext across all $N$.
\paragraph{Takeaways.}
(i) This accuracy experiment isolates transport-fault effects; it does not claim
that the current Refresh skeleton has negligible approximation error. (ii) BCH without interleaving becomes unreliable as $N$ increases (more blocks $\Rightarrow$ higher chance some block exceeds $t$). (iii) Automorphism-based interleaving keeps \emph{expected} accuracy within ${<}0.5\%$ of plaintext across fault rates and ring sizes, consistent with Fig.~\ref{fig:acc-vs-p} and Fig.~\ref{fig:acc-vs-N}.

\section{Conclusion and Future Work}

This paper presented a CKKS-oriented framework for web services that combines
two ideas: a fixed-modulus Refresh-based workflow and a Ring--BCH-inspired
transport-reliability layer. The current prototype should be interpreted as a
step toward this design rather than a complete optimized implementation. In
particular, the Refresh experiment measures an experimental public
homomorphic decrypt-and-reencrypt skeleton using a low-degree surrogate, not a
full CKKS EvalMod or bootstrapping circuit. Separately, the transport-fault
experiments show that BCH coding with interleaving sharply reduces failure under
bursty coefficient faults and keeps modeled downstream accuracy close to the
plaintext baseline. These results support the feasibility of the reliability
layer and identify the main remaining implementation gap: replacing the Refresh
skeleton with a complete and validated CKKS bootstrapping/rounding circuit.

\paragraph{Future Works.}
We plan several directions for optimization and investigation. This includes further optimizing the underlying ring arithmetic using structure-aware techniques~\cite{cheon2023multiprecision}, refining the relinearization process for the Refresh operation, and preprocessing~\cite{zapico2022caulk,zapico2022baloo,bruggemann2023flute}. We will also seek to accelerate the Ring–BCH encoding and decoding with vectorized algorithms and explore potential hardware offloading~\cite{lee2023config}. Finally, we will benchmark our framework against recent CKKS optimizations and investigate the integration of lightweight, TFHE-style bootstrapping under unified security analyses~\cite{micciancio2021bootstrapping,bae2024bits,cheon2025ship, BaePCMM24}.

\bibliographystyle{ACM-Reference-Format} % or plain, ieeetr, etc.
\balance
\bibliography{references}

\appendix
\section{Concrete Algorithms}
% Preamble (once in the paper)
% \usepackage{algorithm}
% \usepackage{algpseudocode}
% \algrenewcommand\algorithmicrequire{\textbf{Input:}}
% \algrenewcommand\algorithmicensure{\textbf{Output:}}
Algorithm~\ref{alg:ring-bch} describes the encoding and decoding process for the Ring BCH layer, and Algorithm~\ref{alg:seg-enc} describes segment encoding and decoding with a BCH code of parameter $(127,101,3)$ and an algebraic structure preserving permutation.

\begin{algorithm}[!t]
\caption{\textsc{RingBCH\_Encode} and \textsc{RingBCH\_Decode}}
\label{alg:ring-bch}
\small
\begin{algorithmic}[1]

\STATE \textbf{Setup:} $R=\mathbb{Z}_{2^k}[X]/(X^N{+}1)$ with $N$ odd; monic $g_k\mid (X^N{+}1)$, $g_k\bmod 2=g_2$; code $\mathcal{C}=(g_k)\subset R$; $K=N-\deg g_k$.

\vspace{0.25em}
\STATE \textbf{Procedure} \textsc{RingBCH\_Encode}$(u)$
\STATE \textbf{Input:} $u\in \mathbb{Z}_{2^k}[X]$, $\deg u < K$
\STATE $w \gets u\,X^{\,N-K}$;\quad $r \gets w \bmod g_k$ in $\mathbb{Z}_{2^k}[X]$
\STATE \RETURN $c \gets (w + r)\bmod (X^N{+}1)\in\mathcal{C}$

\vspace{0.4em}
\STATE \textbf{Procedure} \textsc{RingBCH\_Decode}$(r)$
\STATE \textbf{Input:} $r\in R$ (received)
\STATE $\bar r \gets r \bmod 2$;\quad $S_j \gets \bar r(\bar\alpha^{\,b+j})$ for $j=0,\dots,\delta-2$
\STATE $(\Lambda,\Omega) \gets \textsc{KeyEquationSolve}(S)$;\quad $\mathcal{I} \gets \textsc{ChienSearch}(\Lambda)$
\STATE $e^{(1)} \gets \textsc{Forney}(\Lambda,\Omega)$ \quad \emph{(support $\mathcal{I}$ and mod-2 error values)}
\FOR{$\ell=1$ \TO $k{-}1$}
  \STATE $e^{(2^{\ell+1})} \gets \textsc{LiftDigit}\!\left(r,\ \sum_{j=0}^{\ell-1}2^j e^{(2^{j+1})},\ \mathcal{I}\right)$
\ENDFOR
\STATE $e \gets \sum_{\ell=0}^{k-1}2^\ell e^{(2^{\ell+1})}$;\quad $c \gets r - e$
\STATE $u \gets \textsc{SystematicInverse}(c)$ \quad \emph{(divide by $X^{N-K}$ and reduce mod $g_k$)}
\STATE \RETURN $c$ \textbf{ (and }$u$\textbf{)}
\end{algorithmic}
\end{algorithm}

\begin{algorithm}[!t]
\caption{\textsc{SegmentEncodeInR} and \textsc{DecodeConcatInR} (no external packing)}
\label{alg:seg-enc}
\small
\begin{algorithmic}[1]

\STATE \textbf{Setup:} $n=127$, BCH$(127,101,3)$ lifted to $g_k$; ideal $\mathcal{C}=(g_k)\subset R$; automorphism $\sigma_s$ with $\gcd(s,N)=1$.

\vspace{0.25em}
\STATE \textbf{Procedure} \textsc{SegmentEncodeInR}$(m_{\rm bin}\in\{0,1\}^{M})$
\STATE \textbf{Output:} ciphertexts $\{C^{(i)}\}_{i=0}^{h-1}$
\STATE $h\gets \lceil M/101\rceil$; pad $m_{\rm bin}$ to length $101h$
\FOR{$i=0$ \TO $h-1$}
  \STATE $u^{(i)}\gets m_{\rm bin}[i\cdot 101..(i{+}1)\cdot 101{-}1]$
  \STATE $u^{(i)}(X)\gets \textsc{PlaceLSB}\big(u^{(i)}\big)$ on the $K{=}101$ message positions
  \STATE $c^{(i)}\gets \textsc{RingBCH\_Encode}\!\big(u^{(i)}(X)\big)$
  \STATE $c^{(i)}\gets \sigma_s\!\big(c^{(i)}\big)$ \quad \emph{(code-preserving interleave)}
  \STATE $C^{(i)}\gets \textsc{Enc}\!\big(c^{(i)}\big)$
\ENDFOR
\STATE \RETURN $\{C^{(i)}\}_{i=0}^{h-1}$

\vspace{0.4em}
\STATE \textbf{Procedure} \textsc{DecodeConcatInR}$\big(\{C^{(i)}\}_{i=0}^{h-1}\big)$
\STATE \textbf{Output:} $\hat m_{\rm bin}\in\{0,1\}^{M}$
\FOR{$i=0$ \TO $h-1$}
  \STATE $r^{(i)}\gets \textsc{Dec}\!\big(C^{(i)}\big)$
  \STATE $\hat c^{(i)}\gets \sigma_{s^{-1}}\!\big(r^{(i)}\big)$
  \STATE $\hat u^{(i)}\gets \textsc{RingBCH\_Decode}\!\big(\hat c^{(i)}\big)$
\ENDFOR
\STATE \RETURN $\hat u^{(0)}\parallel\cdots\parallel \hat u^{(h-1)}[0..M{-}1]$

\end{algorithmic}
\end{algorithm}

\section{Correctness of the Non-Leveled Scheme}
\label{subsec:nonleveled-correctness}
\paragraph{Sketch.}
Let $R_q=\mathbb{Z}[X]/(X^N{+}1,q)$ and encode at fixed scale $\Delta$; every ciphertext is an RLWE sample $(b,a)=(m+e-a s,\ a)$ with small noise $e$. Maintain the invariant $\mathcal I(c): \mathrm{scale}(c)=\Delta$ and $\|e(c)\|_\infty\le B$ with $\Delta B\le q/8$. Enc/Dec correctness follows since decryption yields $m+e'$ and, under the bound, rounding in the canonical embedding recovers slots up to standard CKKS approximation. \textsf{Add} preserves scale and sums noises; \textsf{Mult}+relinearization produces pre-refresh noise $B_{\mathrm{mult}}$ that is (standardly) a bilinear-plus-linear function of input bounds and gadget noise and is the only step that can threaten $\Delta B\le q/8$. \textbf{Refresh} homomorphically forms $\mathrm{Enc}(m{+}e)$, applies a low-degree polynomial $R(x)\!\approx\!\mathrm{round}(x/\Delta)$, rescales back to $\Delta$, and re-randomizes; choosing degree/precision and flooding variance ensures a post-refresh bound $B'\!\le\!B_{\mathrm{enc}}$ with $\Delta B'\!\le\!q/8$, restoring $\mathcal I$ independent of prior depth. Because the plaintext Ring–BCH code $(g_k)\!\subset\!R$ is an ideal, $R$-linear evaluation preserves code membership; automorphism interleaving merely permutes coordinates, and BCH decoding corrects whenever each block has $\le t$ errors after transport. Scheduling \textbf{Refresh} whenever $B_{\mathrm{mult}}\!\ge\!B^\star$ yields, by induction over the circuit, that $\mathcal I$ holds at every step and the final slot error is $\varepsilon_{\mathrm{enc}}+n_R\varepsilon_R+\varepsilon_{\mathrm{lin}}$; all ciphertexts remain RLWE samples, so semantic security is unchanged.

\section{Statement and Proof of Lemmas, Propositions, and Theorems}
Assume that a polynomial $a(X) \in R = \mathbb{Z}[X]/(\Phi_M(X))$ is sampled from a uniform distribution or a discrete Gaussian distribution, and its nonzero coefficients are independently and identically distributed. Since $a(\zeta_M)$ is the inner product of the coefficient vector of $a$ and the fixed vector $(1, \zeta_M, \ldots, \zeta_M^{N-1})$, which has Euclidean norm $\sqrt{N}$, the random variable $a(\zeta_M)$ has variance $V = \sigma^2 N$, where $\sigma^2$ is the variance of each coefficient of $a$. Hence, $a(\zeta_M)$ has variances $V_U = q^2 N / 12$, $V_G = \sigma^2 N$, and $V_Z = \rho N$ when $a$ is sampled from $U(R_q)$, $\mathrm{DG}(\sigma^2)$, and $\mathrm{ZO}(\rho)$, respectively. In particular, $a(\zeta_M)$ has variance $V_H = h$ when $a(X)$ is chosen from $\mathrm{HWT}(h)$. Moreover, we can assume that $a(\zeta_M)$ is distributed approximately as a Gaussian random variable over the complex plane, since it is a sum of many independent and identically distributed random variables. All evaluations at roots of unity $\zeta_M^j$ share the same variance. Hence, we adopt $6\sigma$ as a high-probability bound on the canonical embedding norm of $a$ when each coefficient has variance $\sigma^2$. For the product of two independent random variables that are approximately Gaussian with variances $\sigma_1^2$ and $\sigma_2^2$, we use a high-probability bound of $16\sigma_1\sigma_2$.

\subsection{Lemma 1}
\begin{lemma}
Let \( m(X) =Ecd(z; \Delta) \) and \( c =Enc_{\mathsf{pk}}(m(X)) \) be a fresh ciphertext. Let $\epsilon_{enc} = m
(X)-Dec_{sk}(c)$. Then, except with negligible probability, $
\left\|\epsilon_{enc} \right\|_{can,\infty} \leq B_{enc}
=
8\sqrt{2}\,\sigma N + 6\sigma\sqrt{N} + 16\sigma\sqrt{hN}$. Furthermore, decoding is robust given \( \Delta > N + 2 B_{enc}\). By construction, $B_{refresh} \gtrsim B_{enc}$.
\end{lemma}
\begin{proof}
The encryption noise bound $B_{enc}$ is computed via the following inequality:
\[
    \big\| \langle c, \mathsf{sk} \rangle - m \mod q_L \big\|_{can, \infty}
    = \big\| v \cdot e + e_0 + e_1 \cdot \mathsf{sk} \big\|_{can, \infty}.
\]
We apply the triangle inequality to obtain:
\[
    \leq \| v \cdot e \|_{can, \infty}
    + \| e_0 \|_{can, \infty}
    + \| e_1 \cdot \mathsf{sk} \|_{can, \infty}.
\]
Each term is bounded individually as follows:
\begin{align*}
\| v \cdot e \|_{can, \infty} \leq 8\sqrt{2} \cdot \sigma N, \quad
    \| e_0 \|_{can, \infty} \leq 6\sigma \sqrt{N}, \quad\\
    \| e_1 \cdot \mathsf{sk} \|_{can, \infty} \leq 16\sigma \sqrt{hN}.    
\end{align*}
Thus, the total encryption noise satisfies:
\[
    B_{\text{clean}} \leq 8\sqrt{2} \cdot \sigma N + 6\sigma \sqrt{N} + 16\sigma \sqrt{hN}.
\]
\end{proof}
\subsection{Lemma 2}
\begin{lemma}\label{lem:enc-noise}
Let $z$ be a plaintext vector and let
$\Delta$ be the scaling factor. Define $
\widetilde m = \Delta\bigl(\pi^{-1}(z)\bigr)$, $\widehat m = \bigl\lfloor\widetilde m\bigr\rceil\in R$ with 
the rounding error $\epsilon_{\mathrm{ecd}} = m - \widetilde m$.
We have: 
$||\epsilon_{\mathrm{ecd}}||_{can,\infty}\leq B_{ecd}=
\frac{\|\sigma_1\|_{\mathrm{op}}}{2\Delta}$,
where $\|\sigma_1\|_{\mathrm{op}}\!\approx\!\sqrt{N}$.
\end{lemma}
\begin{proof}
Each coefficient of $\widetilde m=\Delta\,\pi^{-1}(z)$ is a real number.
Rounding coordinate-wise yields
$\widehat m=\widetilde m+\epsilon_R$ with 
$\epsilon_R:=\widehat m-\widetilde m\in R$ and
$\|\epsilon_R\|_\infty\le \tfrac12$. Because the coefficients were scaled by~$\Delta$, we have
$\|\,\epsilon_R\|_\infty\le \tfrac1{2\Delta}$.
By definition $p(\epsilon_{\mathrm{enc}})=\epsilon_R$, so
\[
\|p(\epsilon_{\mathrm{enc}})\|_\infty \;\le\; \frac1{2\Delta}.
\]
Finally, applying the linear map $\sigma_1$ multiplies any vector norm by
at most $\|\sigma_1\|_{\mathrm{op}}$, hence
\[
\bigl\|\sigma_1\!\bigl(\,p(\epsilon_{\mathrm{enc}})\bigr)\bigr\|_\infty
\;\le\;
\frac{\|\sigma_1\|_{\mathrm{op}}}{2\Delta}.
\]
\end{proof}
\subsection{Lemma 3}
\begin{lemma}\label{multi}
Let $(c_i,\Delta_i,B_i)$ encrypt $m_i\in R$ for $i\in\{1,2\}$.
Let $\nu_i:=\|m_i\|_{\mathrm{can},\infty}$ and set $\nu:=\max(\nu_1,\nu_2)$. $\widetilde c:=\textsc{MultConst}\!\big(\tfrac{\Delta}{\Delta_1\Delta_2},\,c_{\mathrm{mult}}\big)$. Then, except with negligible probability, for the common fixed--scale case $\Delta_1=\Delta_2=\Delta$, we have $\|\mathrm{Dec}(\widetilde c)-m_1 m_2\|_{\mathrm{can},\infty}
\ \le B_{mult}$ and 
\begin{align*}
B_{mult} =\frac{\nu(B_2+ B_1)+B_1B_2+16\sqrt{3}\sigma^2 N}{\Delta}. 
\end{align*}
\end{lemma}
\begin{proof}
Write $\mathrm{Dec}(c_i)=m_i+e_i$ so $\|e_i\|_{\mathrm{can},\infty}\le B_i$.
Before relinearization,
\[
(b_1+a_1 s)(b_2+a_2 s)=d_0+d_1 s+d_2 s^2.
\]
Adding $d_2\cdot\mathsf{evk}$ replaces $d_2 s^2$ by $d_2(s^2+e_{\mathsf{evk}})$, so the decrypted value of $c_{\mathrm{mult}}$ equals
\(
(m_1+e_1)(m_2+e_2)+d_2 e_{\mathsf{evk}}.
\)
Hence, the pre-scaling error is
\(
e_{\mathrm{mult}}=m_1 e_2+m_2 e_1+e_1 e_2+d_2 e_{\mathsf{evk}}.
\)
By submultiplicativity in the canonical embedding,
$\|m_1 e_2\|\le \nu_1 B_2$, $\|m_2 e_1\|\le \nu_2 B_1$, $\|e_1 e_2\|\le B_1 B_2$ and $\|d_2 e_{\mathsf{evk}}\|\le \|d_2\|\,\|e_{\mathsf{evk}}\|=U_{\times} B_{\mathsf{evk}}$.
Finally, multiplying the ciphertext by $\Delta/(\Delta_1\Delta_2)$ scales the error by the same factor. Using the assumption for concrete bounds as shown in the lemma statement.
\end{proof}

\subsection{Lemma 4}
\begin{lemma}
Let $c_1, c_2$ be ciphertexts encrypting $m_1, m_2 \in R$ with decryption noise bounded by $B_1$ and $B_2$ of the same scale $\Delta$. Let $c_{\mathrm{add}} = c_1 + c_2$. Then: $\| \epsilon_{\mathrm{add}} \|_{can,\infty} \leq B_{add}= B_1 + B_2$. 
\end{lemma}
\begin{proof}
Homomorphic addition of ciphertexts is component-wise: $c_{\mathrm{add}} = c_1 + c_2$. The decryption of $c_{\mathrm{add}}$ yields: $\mathrm{Dec}(c_{\mathrm{add}}) = m_1 + m_2 + (\epsilon_1 + \epsilon_2)$. Therefore:
\[
\| \epsilon_{\mathrm{add}} \|_\infty \leq \| \epsilon_1 \|_\infty + \| \epsilon_2 \|_\infty \leq B_1 + B_2.
\]
Correctness holds as long as $\Delta > 2 \cdot \| \epsilon_{\mathrm{add}} \|_\infty$.
\end{proof}
\subsection{Lemma 5}
\begin{lemma}[Addition and Multiplication by Constant]
Let \( (c, B_c) \) be a CKKS encryption of \( m \). For a constant \( a \in R\), $c_a := c + (a,0)$, $c_m := a \cdot c$, 
where the ciphertexts \( (c_a, B_c) \) and \( (c_m, \|a\|_{can,\infty} \cdot B_c) \) are valid encryptions of \( m + a \) and \( a \cdot m \), respectively.
\end{lemma}
\begin{proof}
By correctness of the proposed framework, the inner product of \( c \) and the secret key yields \( m + e \) for some error polynomial \( e \) with \( \|e\|_{can,\infty} \leq B \). For the addition case, since the encoded constant \( a \) contributes no noise, the resulting error remains bounded by \( B \). For multiplication, the ciphertext becomes \( a \cdot (m + e) = a m + a e \), which implies that the new error is bounded by \( \|a\|_{can,\infty} \cdot B \).
\end{proof}
\subsection{Lemma 6}
\begin{lemma}[Polynomial Evaluation]
\label{lem:poly-adaptive}
Let $f(x)=\sum_{j=0}^{d} a_j x^{\,j}$ with real coefficients
$a_j$, and let $(c_0,B_0)$ be a valid ciphertext
encrypting $m\in R$ with initial coefficient-error bound
$B_0=B_{\mathrm{enc}}$. Fix a refresh threshold $B^{*}$.
The extended Algorithm~\ref{poly} outputs a ciphertext $(c_f,B_f)$ such that $c_f = Enc(f(m))$ with $B_f\approx \min\{\sum^d_{j=0}|a_j|B_{enc},\;B^*\}$.
\end{lemma}

\begin{proof}
All norms are $\|\cdot\|_\infty$. One squaring (multiply, relinearize, exact rescale) of an encryption of $m$ with error $e$ gives
\[
Dec(\mathbf{Mult}(c,c))/\Delta-m^2 \;=\; m e + e m + e^2 + e_{\mathrm{ks}},
\]
so with $\nu=\|m\|_\infty$ and $C_{\mathrm{ks}}=16\sqrt{3}\sigma^2N$,
\[
B_{\mathrm{sq}}(B)\;\le\;\frac{2\nu B + B^2 + C_{\mathrm{ks}}}{\Delta}.
\]
The algorithm sets the new error to $\min\{B_{\mathrm{sq}}(B),\,B^*\}$, and calls \textbf{Refresh} whenever needed to clamp it to $B_{\mathrm{enc}}$. Hence after any powering chain (binary powering for $m^j$), the monomial ciphertext has error $\le \min\{G(B_{\mathrm{enc}}),B^*\}\le B^*$, and if a refresh occurred then $\le B_{\mathrm{enc}}$.

Plaintext scaling by $a_j$ multiplies the error by $|a_j|$. Summing monomials adds errors. Execute the final sum without refresh while the running bound stays $\le B^*$; if it would exceed $B^*$, refresh once to $B_{\mathrm{enc}}$ and continue. Consequently,
\[
B_f \;\le\; \min\Big\{\sum_{j=0}^d |a_j|\,B_{\mathrm{enc}},\; B^*\Big\}.
\]
All bounds hold except with negligible probability from subgaussian tails of encryption and EVK noise.
\end{proof}

\subsection{Proof of Proposition~\ref{pro:bin_vs_std}}
\label{proof_binvsstd}
\begin{proof}
Followed by simplifying the multiplication noise equations.
\end{proof}

\subsection{Proof of Theorem~\ref{thm_analy}}
\label{pf_analy}
\begin{proof}[Proof Sketch]
Since $f(x)$ is analytic, it equals its power series expansion in some radius of convergence around 0. We choose $D$ such that the omitted tail $\sum_{j>D} a_j m^j$ is at most $\epsilon$ in magnitude (by an appropriate form of Taylor’s remainder bound). The truncated polynomial $P_D(x) = \sum_{j=0}^{D} a_j x^j$ is then evaluated homomorphically as described. Each exponentiation $x^j$ is performed by repeated squaring with refresh: after each multiplication, the noise would grow to roughly $(1+h)B^2$ (dominated by the quadratic term in $B$), so we refresh whenever the noise bound would exceed $B^*$. This ensures that each power $x^j$ is obtained with noise $\approx B_{\text{enc}}$ (by Lemma 6). Multiplying by the plaintext coefficient $a_j$ does not increase the noise. Summing the encrypted terms contributes additive noise up to $\sum_{j=0}^D |a_j|\, B_{\text{enc}}$. Finally, because we cap the noise at $B^*$ via refresh, the output noise $B_f$ is at most $B^*$. In summary, \[ B_f \;\approx\; \min\Big\{\, \sum_{j=0}^{D} |a_j|\, B_{\text{enc}},\; B^* \Big\}\,, \] and including the truncation error $\epsilon$ from ignoring high-degree terms, the total error is bounded by $\epsilon$ as required.
\end{proof}
\paragraph{Note:} Given Theorem~\ref{thm_analy} and Lemma~\ref{lem:poly-adaptive}, we compare the accuracy of polynomial evaluation under standard CKKS and our non-leveled CKKS variant by analyzing the resulting \emph{relative error}. Let \(f(x)=\sum_{j=0}^{d} a_j x^j\) be a polynomial with coefficients \(a_j\in\mathbb{R}\). Recall the relative error of a ciphertext as \(\beta := B_f/\Delta\), where \(B_f\) is the ciphertext’s absolute error and \(\Delta\) is the plaintext scaling factor.

In standard CKKS, the evaluated ciphertext is typically tracked as
\((c',\,\ell-\lceil\log_2 d\rceil,\,M_f,\,\beta_d M_f)\), and its relative noise satisfies
\(\beta_d \le d\,\beta_0 + (d-1)\beta^*\), where \(\beta_0\) is the initial relative error and \(\beta^*\) is the per-level noise introduced by rescaling.

In our variant, the resulting ciphertext has absolute noise
\[
B_f=\min\left\{\sum_{j=0}^{d} |a_j|\,B_{\mathrm{enc}},\; B^*\right\},
\]
and hence the relative error is
\[
\beta_f=\min\left\{\frac{\sum_{j=0}^{d} |a_j|\,B_{\mathrm{enc}}}{\Delta},\; \beta^*\right\}.
\]
In the unsaturated regime \(\sum_{j=0}^{d} |a_j|\,B_{\mathrm{enc}}\le B^*\), we compare relative errors by checking when
\[
\frac{\sum_{j=0}^{d} |a_j|\,B_{\mathrm{enc}}}{\Delta}
\;<\;
d\,\beta_0 + (d-1)\beta^*.
\]

\subsection{Proof of Theorem~\ref{thm:indcpa}}
\label{pf_indcpa}

\begin{proof}
We prove security by a sequence of games. Write the public key as
\[
  \mathsf{pk}=(b,a),\qquad b=-a s+e,
\]
where \(a\leftarrow R_q\), \(s\leftarrow D_s\), and \(e\leftarrow \chi\).
An encryption of \(m\in R_q\) has the form
\[
  (c_0,c_1)=(v b+m+e_0,\; v a+e_1),
\]
where \(v\leftarrow D_v\) and \(e_0,e_1\leftarrow \chi\).

\paragraph{Game \(G_0\).}
This is the real IND-CPA experiment. The adversary receives
\(\mathsf{pk}=(b,a)\), submits two equal-length messages \(m_0,m_1\),
and receives
\[
  c^\star=(v b+m_\beta+e_0,\; v a+e_1)
\]
for a uniform challenge bit \(\beta\in\{0,1\}\).

\paragraph{Game \(G_1\).}
This game is identical to \(G_0\), except that the public key is replaced
by a uniformly random pair
\[
  (b,a)\leftarrow R_q^2.
\]
By decisional RLWE with secret distribution \(D_s\), the real public key
\((b,a)=(-as+e,a)\) is computationally indistinguishable from uniform.
Therefore,
\[
  \left|\Pr[G_0\Rightarrow 1]-\Pr[G_1\Rightarrow 1]\right|
  \leq
   Adv ^{\mathrm{RLWE}}_{R_q,D_s,\chi}(\mathcal B_1)
\]
for some PPT distinguisher \(\mathcal B_1\).

\paragraph{Game \(G_2\).}
This game is identical to \(G_1\), except that the challenge ciphertext
mask
\[
  (v b+e_0,\; v a+e_1)
\]
is replaced by a uniformly random pair
\[
  (r_0,r_1)\leftarrow R_q^2.
\]
Since in \(G_1\) the elements \(a,b\) are uniform and independent, the
pair
\[
  (b,\; vb+e_0),\qquad (a,\; va+e_1)
\]
is a two-sample RLWE instance with secret \(v\leftarrow D_v\), up to the
ordering of the coordinates. Hence, by the multi-sample decisional RLWE
assumption for secret distribution \(D_v\),
\[
  \left|\Pr[G_1\Rightarrow 1]-\Pr[G_2\Rightarrow 1]\right|
  \leq
   Adv ^{\mathrm{RLWE}}_{R_q,D_v,\chi,2}(\mathcal B_2)
\]
for some PPT distinguisher \(\mathcal B_2\).

In \(G_2\), the challenge ciphertext is
\[
  c^\star=(r_0+m_\beta,\; r_1),
\]
where \((r_0,r_1)\) is uniform over \(R_q^2\). Since adding a fixed
message \(m_\beta\) to a uniform ring element preserves uniformity, the
distribution of \(c^\star\) is independent of \(\beta\). Therefore
\[
  \Pr[G_2\Rightarrow 1]=\frac12.
\]

Combining the hybrids gives
\[
 Adv _{\Pi}^{\mathsf{ind\mbox{-}cpa}}(\mathcal A)
\leq
 Adv ^{\mathrm{RLWE}}_{R_q,D_s,\chi}(\mathcal B_1)
+
 Adv ^{\mathrm{RLWE}}_{R_q,D_v,\chi,2}(\mathcal B_2).
\]
Both terms are negligible by assumption, so the encryption layer is
IND-CPA secure.
\end{proof}

\subsection{Proof of Theorem~\ref{thm_indcpa_bch}}
\label{pf_indcpa_bch}
\begin{proof}
Given an adversary \(\mathcal A\) against \(\mathcal S^\star\), construct
an adversary \(\mathcal B\) against \(\mathcal S\). The adversary
\(\mathcal B\) forwards the public key it receives to \(\mathcal A\).
When \(\mathcal A\) outputs challenge messages \(m_0,m_1\), \(\mathcal B\)
computes
\[
  M_0=\mathsf{BCHEnc}(m_0),\qquad
  M_1=\mathsf{BCHEnc}(m_1),
\]
and submits \(M_0,M_1\) to its own IND-CPA challenger for \(\mathcal S\).
The challenger returns an encryption of \(M_\beta\), which \(\mathcal B\)
forwards unchanged to \(\mathcal A\). Finally, \(\mathcal B\) outputs
whatever bit \(\mathcal A\) outputs.

Because the Ring--BCH encoder is public and deterministic, the view of
\(\mathcal A\) in this simulation is exactly its view in the real
IND-CPA experiment for \(\mathcal S^\star\). Therefore the two advantages
are equal up to the advantage of \(\mathcal B\) against the base scheme.
Thus, if \(\mathcal S\) is IND-CPA secure, so is \(\mathcal S^\star\).
\end{proof}
\end{document}